\documentclass[iop,apj]{emulateapj}

\usepackage{graphicx,amsmath,amssymb}
\usepackage{multirow}

\shorttitle{The \emph{Swift} GRB Redshift Distribution}
\shortauthors{Jakobsson et al.}

\begin{document}





\title{THE OPTICALLY UNBIASED GRB HOST (TOUGH) SURVEY. \\ 
III. REDSHIFT DISTRIBUTION}


\author{P.~Jakobsson\altaffilmark{1}, J.~Hjorth\altaffilmark{2},
D.~Malesani\altaffilmark{2}, R.~Chapman\altaffilmark{1,3}, 
J.~P.~U.~Fynbo\altaffilmark{2}, N.~R. Tanvir\altaffilmark{4},
B.~Milvang-Jensen\altaffilmark{2}, P.~M.~Vreeswijk\altaffilmark{1}, 
G. Letawe\altaffilmark{5}, and R.~L.~C.~Starling\altaffilmark{4}}


\altaffiltext{1}{Centre for Astrophysics and Cosmology, Science Institute, 
University of Iceland, Dunhagi 5, 107 Reykjav\'ik, Iceland}
\altaffiltext{2}{Dark Cosmology Centre, Niels Bohr Institute, University of 
Copenhagen, Juliane Maries Vej 30, 2100 Copenhagen \O, Denmark}
\altaffiltext{3}{Centre for Astrophysics Research, University of 
Hertfordshire, Hatfield, Herts AL10 9AB, UK}
\altaffiltext{4}{Department of Physics and Astronomy, University of Leicester, 
University Road, Leicester, LE1 7RH, UK}
\altaffiltext{5}{D\'epartement d'Astrophysique, G\'eophysique et 
Oc\'eanographie, ULg, All\'ee du 6 ao\^{u}t, 17 - B\^{a}t. B5c B-4000 
Li\`ege (Sart-Tilman), Belgium}



\begin{abstract}
We present 10 new gamma-ray burst (GRB) redshifts and another five redshift 
limits based on host galaxy spectroscopy obtained as part of a large program
conducted at the Very Large Telescope (VLT). The redshifts span the range 
$0.345 \leq z \lesssim 2.54$. Three of our measurements revise incorrect 
values from the literature. The homogeneous host sample researched here 
consists of 69 hosts that originally had a redshift completeness of 55\% (with 
38 out of 69 hosts having redshifts considered secure). Our project, including 
VLT/X-shooter observations reported elsewhere, increases this fraction to 77\%
(53/69), making the survey the most comprehensive in terms of redshift 
completeness of any sample to the full \emph{Swift} depth, analyzed to date.
We present the cumulative redshift distribution and derive a conservative, 
yet small, associated uncertainty. We constrain the fraction of \emph{Swift} 
GRBs at high redshift to a maximum of 14\% (5\%) for $z > 6$ ($z > 7$).
The mean redshift of the host sample is assessed to be $\langle z \rangle
\gtrsim 2.2$, with the 10 new redshifts reducing it significantly. 
Using this more complete sample, we confirm previous findings that the
GRB rate at high redshift ($z\gtrsim3$) appears to be in excess of
predictions based on assumptions that it should follow conventional 
determinations of the star formation history of the universe, combined with 
an estimate of its likely metallicity dependence. This suggests that either 
star formation at high redshifts has been significantly underestimated, for 
example due to a dominant contribution from faint, undetected galaxies, or 
that GRB production is enhanced in the conditions of early star formation, 
beyond that usually ascribed to lower metallicity.
\end{abstract}





\section{INTRODUCTION}

Determining the statistical properties of gamma-ray bursts (GRBs) has long 
been compromised by inhomogeneous selection and a bias against optically 
dark bursts.  With {\it Swift} \citep{gehrels} it has become possible to 
construct much more uniform samples, and to target the host galaxies even of 
optically faint bursts via X-Ray Telescope (XRT) localizations, for which 
redshifts could not be determined from the afterglows.
\par
We have been securing GRB host galaxy information for a homogeneous sample of 
69 \emph{Swift} GRBs with a large program at the Very Large Telescope (VLT). 
The first observations of The Optically Unbiased GRB Host (TOUGH) sample
were obtained on 2006 February 23 with the survey concluding on 2008 August 
29. The immediate goals are to determine the host luminosity function (LF), 
study the effects of reddening, determine the fraction of Ly$\alpha$ emitters 
in the hosts, and obtain redshifts for targets without a reported one. The 
sample has been carefully selected and obeys strict and well-defined criteria. 
To optimize the survey, we focused on systems with the best observability, 
which also have the best available information. The survey design and 
catalogs are presented in \citet{jensLP}, the fundamental properties of the 
hosts in \citet{danieleLP}, the Ly$\alpha$ emission in \citet{boLP}, and
new VLT/X-shooter redshifts in \citet{thomas}.
\par 
This paper presents the first TOUGH campaign for missing redshifts. We 
attempted spectroscopic observations of most TOUGH host candidates with 
$R \lesssim 25$\,mag that did not have a reported reliable redshift. A total 
of 19 candidates were spectroscopically observed for an aggregate of nearly 
30\,hr with the aim of acquiring redshift information. In addition, we include 
a new redshift measurement for GRB\,060908 reported in \citet[][see also
\citealt{johan_sample}]{boLP}. The details of our observations and reductions 
are described in the next section. New redshifts and redshift limits for each 
observed host are presented in \S \ref{z.sec}. We model the redshift 
distribution and compare it to the observed one in \S \ref{dist.sec}. Finally, 
the implications of our observational efforts are discussed in 
\S \ref{dis.sec}. We adopt a cosmology where the Hubble parameter is 
$H_0 = 70.4$\,km\,s$^{-1}$\,Mpc$^{-1}$, $\Omega_{\mathrm{m}} = 0.27$ and
$\Omega_{\Lambda} = 0.73$ \citep{jarosik}.
\par
The reduced data from this work will be available from 
ESO\footnote{\texttt{http://archive.eso.org}} and from the TOUGH 
Web site.\footnote{\texttt{http://www.dark-cosmology.dk/TOUGH}}

\section{OBSERVATIONS AND DATA REDUCTION}

The host spectroscopic observations were carried out between 2006 May 30
and 2008 August 29. The FORS2 instrument was used as well as FORS1 
\citep{appen} before and after the blue CCD upgrade. Four different grism 
setups were mainly used: 300V, 300V+GG375, 600z+OG590, and 
600RI+GG435.\footnote{See 
\texttt{http://www.eso.org/sci/facilities/paranal/ \\ instruments/fors/doc/} 
for more details.} We requested a seeing constraint of 1\farcs2 for all 
observations apart from GRB\,060923C for which 0\farcs8 was required (\S 
\ref{060923C.sec}). For the majority of the observations, this goal was 
accomplished. More detailed information is listed in Table~\ref{z.tab}.
\par
The data reduction was performed following standard techniques for bias and 
flat-field corrections. The individual spectra were then cosmic ray cleaned 
using the method of \cite{vanD}. If the host trace was clearly visible
in individual images, the spectra were optimally extracted for each
two-dimensional image separately. Otherwise, the two-dimensional images
were aligned, averaged, and the host spectrum extracted if detected.
\par
The wavelength calibration was applied using a He\-NeAr lamp spectrum obtained 
in the morning after the science observations. The root mean square scatter
around the wavelength calibration fit was roughly 0.2\,\AA\ for the 300V grism 
and less than 0.1\,\AA\ for the higher resolution grisms. The individual 
wavelength-calibrated spectra were averaged, and the corresponding Poisson 
error spectra (calculated by the IRAF/apall task) were quadratically averaged. 
Finally, flux calibration was applied by using standard star observations 
carried out every night a host was observed. We note, however, that the 
transparency constraint was set to ``thin cirrus'' implying that absolute
flux calibration should be interpreted with caution.

\section{NEW REDSHIFTS AND LIMITS}

\label{z.sec}

A total of 20 host systems were observed (including GRB\,060908), of which 
only 10 had a reported optical afterglow (OA) or near-infrared afterglow 
(NIRA). In only two cases (GRBs\,060805A and 070808) there is more than one 
galaxy detected within or on the border of the XRT error circle. For the other
eight systems without an OA/NIRA the host identity is nearly unambiguous with 
the probability of chance projection being fairly low \citep[frequently less 
than 5\%;][]{danieleLP}. Below we discuss each system in detail and justify our 
redshift measurements and limits. When no spectral features are detected we set 
redshift limits in the following way:

\paragraph{} 
If the continuum is visible we estimate, using the error spectrum,
the lowest wavelength at which it is significantly detected (at a significance
level of around 3$\sigma$). The lack of Ly$\alpha$ forest lines is then used 
to set a firm redshift upper limit.

\paragraph{} 
To get a redshift lower limit we note that GRB hosts are in general 
star-forming galaxies and thus display the usual emission lines, such as 
[\ion{O}{2}] $\lambda 3727$, H$\beta$, [\ion{O}{3}] $\lambda 5007$, and
H$\alpha$; see e.g.\ the GHostS 
Web site,\footnote{\texttt{http://www.grbhosts.org/}} Table 3 in 
\citet{savaglio}, and \citet{levesque}. In particular, the non-detection of 
[\ion{O}{2}] $\lambda 3727$ in our 300V spectra immediately sets a redshift 
limit of $z > 0.9$. This value is based on 7200\,\AA, since above it there 
are strong skylines which leave significant residuals in the reduced spectra.

\paragraph{} 
If neither emission lines nor a continuum are detected we follow \citet{grupe} 
and assign a redshift upper limit to bursts with excess (above Galactic) X-ray 
absorbing column density\footnote{We use the afterglow-only late spectrum 
(photon counting mode) obtained from 
\texttt{http://www.swift.ac.uk/xrt\_spectra/}}, calculated at $z = 0$, above 
an equivalent hydrogen column density of $2 \times 10^{21}$\,cm$^{-2}$ 
(including the uncertainty).\footnote{For comparison, one of the sample 
selection criteria, as defined in \citet{jensLP}, is that the Galactic 
extinction is $A_V \le 0.5$\,mag, which corresponds to a hydrogen column 
density of $0.9 \times 10^{21}$\,cm$^{-2}$ according to the relation derived by 
\citet{pre}.} \citet{grupe} used an upper limit of $z = 2$, but we 
will be slightly more conservative and assign an upper limit of $z = 3.5$ to 
these bursts \citep[see Equation~(1) in][]{grupe}. The corresponding rest-frame 
column density is $10^{23}$\,cm$^{-2}$. In the sample of \citet[][see their 
Figure~2]{campana}, no burst exceeds this number, lending support to our choice 
of $z \lesssim 3.5$. This approach was utilized for six events in the whole
TOUGH sample. Among those, five have an $R$-band host detection, implying that
their redshifts have to be $z \lesssim 6$.

\subsection{GRB\,050714B (No OA/NIRA)}

There is a single faint source detected inside the XRT error circle, with a 
bright source close to the southern edge of the error circle. Our 
spectroscopic observations show the brighter one to be an M star. Neither 
emission lines nor a continuum are detected from the fainter object and hence 
no redshift information can currently be obtained for this host candidate. 
However, we conclude that $z \lesssim 3.5$ based on the excess column density 
detected in the X-ray spectrum.

\subsection{GRB\,050822 (No OA/NIRA)}

\begin{figure*}
\epsscale{1.17}
\plottwo{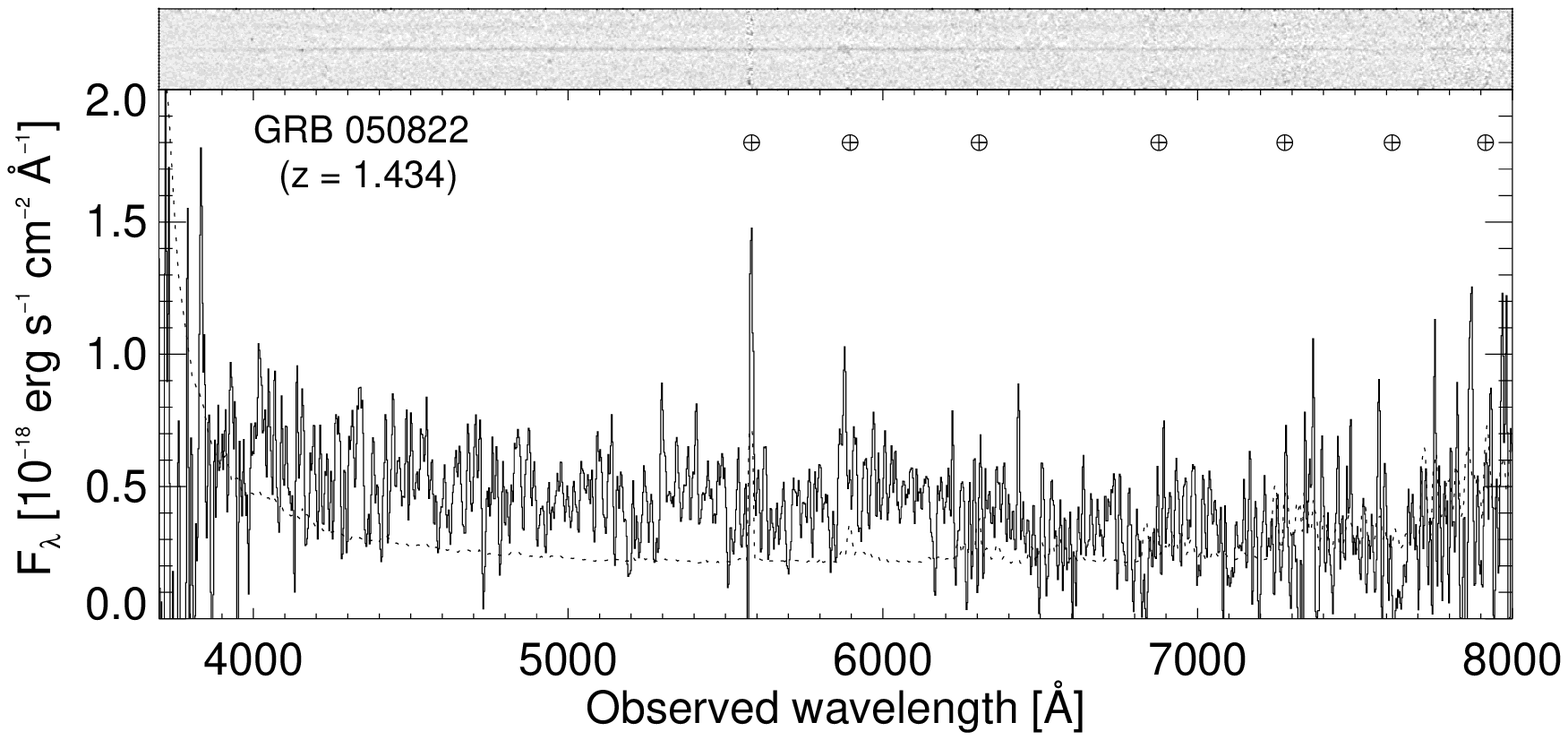}{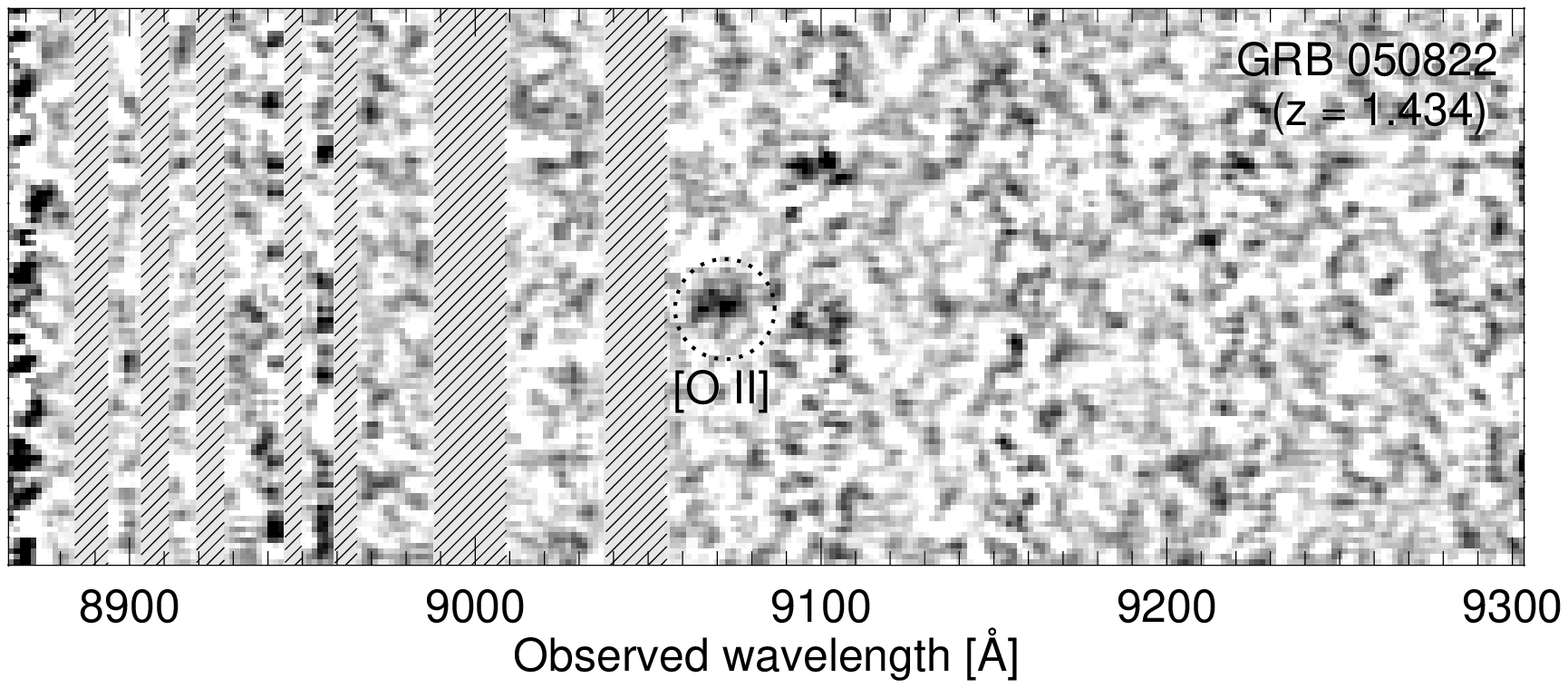}
\caption{One- and two-dimensional spectra (left: 300V; right: 600z) of 
the GRB\,050822 host. There are no emission lines identified in 300V. Telluric 
features and skyline residuals are marked with $\oplus$, whereas the error 
spectrum is plotted as a dotted line. The 600z emission line candidate is 
marked with a circle in the right panel. Hashed regions mark areas strongly 
affected by skylines.}
\label{050822.fig}
\end{figure*}

In the 300V spectrum, the continuum is clearly detected in the 3900--7200\,\AA\ 
region corresponding to the redshift range $0.9 \lesssim z \lesssim 2.2$ (left 
panel of Fig.~\ref{050822.fig}). There is an emission line candidate 
(7$\sigma$) at 9071\,\AA\ (right panel of Fig.~\ref{050822.fig}) that is  most 
likely the [\ion{O}{2}] $\lambda 3727$ doublet at $z = 1.434$. In support of 
this, we note that the spectral extent of the line is significantly broader 
than the extent of emission lines visible in other traces in the 
two-dimensional spectrum.
\par
At this redshift, our spectra would not cover H$\beta$, [\ion{O}{3}] $\lambda 
5007$, and H$\alpha$. Other potential line identifications (of 9071\,\AA) such 
as H$\beta$, [\ion{O}{3}] $\lambda 5007$, and H$\alpha$ are rejected on 
account that we would expect to see additional and roughly similarly bright 
lines in the wavelength range covered.

\subsection{GRB\,050915A (NIRA)}

\begin{figure}
\epsscale{1.17}
\plotone{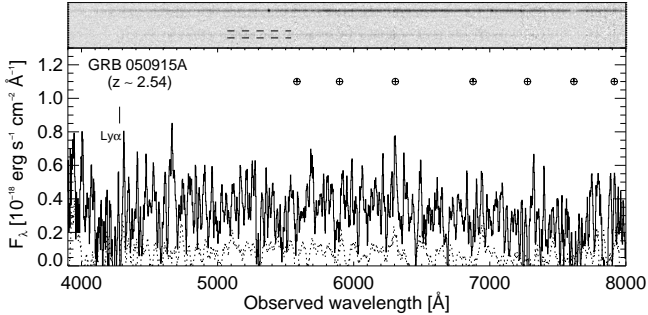}
\caption{One- and two-dimensional spectra (300V) of the GRB\,050915A 
host. In the upper panel, it is the bottom trace between the two horizontal 
dashed lines. The Ly$\alpha$ absorption feature is marked with a 
vertical line, whereas telluric features and skyline residuals are marked 
with $\oplus$. The error spectrum is plotted as a dotted line.}
\label{050915A.fig}
\end{figure}

The host galaxy of the dark \citep{pallidark} GRB\,050915A was previously 
reported in \citet{ovald} and \citet{perley}. In 300V the continuum is clearly 
detected in the 4300--7200\,\AA\ region without unambiguous features, 
corresponding to $0.9 \lesssim z \lesssim 2.5$ (Fig.~\ref{050915A.fig}).
\par
In the upper panel of Fig.~\ref{050915A.fig}, the continuum of another
brighter galaxy on the slit is visible ($z = 0.444$).\footnote{This value
was erroneously reported as the GRB\,050915A host redshift in 
\citet{palliAIP,palliAdSpR,palliAN,palliUSA}.} It clearly contains flux 
down to around 3600\,\AA. Therefore, a spectral break must be present in the 
GRB\,050915A host continuum, which we interpret as the Ly$\alpha$ break. 
More specifically, the 1D spectrum shows a flux drop around 4300\,\AA\ 
corresponding to $z \approx 2.54$. Recent observations by VLT/X-shooter 
confirm this interpretation where we detect [\ion{O}{2}] $\lambda 3727$,
H$\beta$, and [\ion{O}{3}] $\lambda \lambda 4959, 5007$ in emission at a
similar redshift \citep{thomas}.

\subsection{GRB\,051001 (No OA/NIRA)}

\begin{figure}
\epsscale{1.17}
\plotone{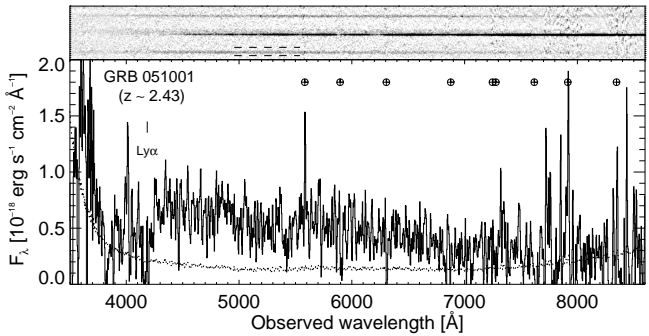}
\caption{One- and two-dimensional spectra (300V) of the GRB\,051001 host.
In the upper panel, it is the bottom trace between the two horizontal dashed 
lines. The Ly$\alpha$ absorption feature is marked with a vertical line, 
whereas telluric features and skyline residuals are marked with $\oplus$. The 
error spectrum is plotted as a dotted line.}
\label{051001.fig}
\end{figure}

The continuum is extremely faint in the 600z spectrum and there are no signs 
of any emission lines. In 300V the continuum is clearly detected in the 
4200--7200\,\AA\ region without unambiguous features, corresponding to 
$0.9 \lesssim z \lesssim 2.5$ (Fig.~\ref{051001.fig}). Thus, GRB\,051001 can 
safely be ruled out as a high-redshift burst as suggested by \citet{salvaZ}.
\par
The 1D spectrum shows a flux drop around 4170\,\AA\ which we interpret as the 
Ly$\alpha$ break at $z \approx 2.43$. Recent observations by VLT/X-shooter 
confirm this interpretation where we detect H$\beta$, [\ion{O}{3}] $\lambda 
5007$, and H$\alpha$ in emission at a similar redshift \citep{thomas}.

\subsection{GRB\,051006 (No OA/NIRA)}

\begin{figure}
\epsscale{1.17}
\plotone{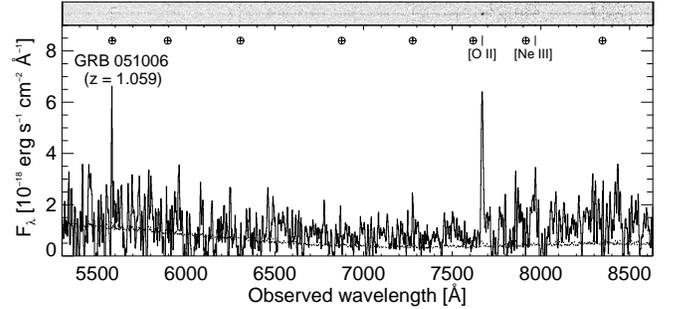}
\caption{One- and two-dimensional spectra (600RI) of the GRB\,051006 
host. Emission lines are marked with vertical lines, whereas telluric 
features and skyline residuals are marked with $\oplus$. The error spectrum is 
plotted as a dotted line.}
\label{051006.fig}
\end{figure}

The 600RI spectrum clearly shows a strong and broad emission line which we 
identify as the [\ion{O}{2}] $\lambda 3727$ doublet at $z = 1.059$. At that 
redshift, we also detect a much weaker line, [\ion{Ne}{3}] $\lambda 3869$ 
(Fig.~\ref{051006.fig}). The stronger line is unlikely to be H$\beta$, 
[\ion{O}{3}] $\lambda 5007$, or H$\alpha$ since additional and roughly 
similarly bright lines would be expected in the wavelength range covered.

\subsection{GRB\,051117B (No OA/NIRA)}

\begin{figure}
\epsscale{1.17}
\plotone{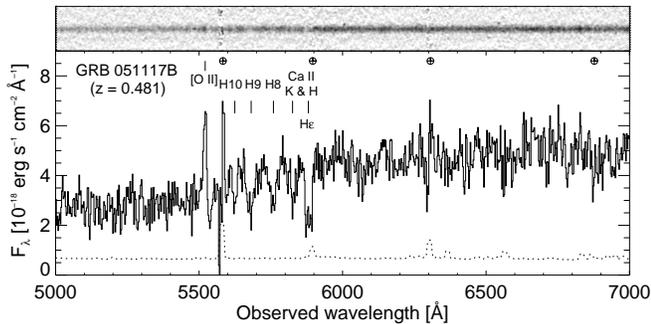}
\caption{One- and two-dimensional spectra (300V) of the GRB\,051117B host. 
Line features are marked with vertical lines, whereas telluric features and 
skyline residuals are marked with $\oplus$. The error spectrum is plotted as 
a dotted line.}
\label{051117B.fig}
\end{figure}

Our 300V spectrum clearly shows a strong emission line in addition to
a few absorption features (Fig.~\ref{051117B.fig}). These features are 
consistent with being [\ion{O}{2}] $\lambda 3727$, H10, H9, H8, H$\varepsilon$
and \ion{Ca}{2} K and H at a common redshift of $z = 0.481$. We note that
H$\varepsilon$ and \ion{Ca}{2} H are blended with a strong skyline residual.

\begin{figure}
\epsscale{1.17}
\plotone{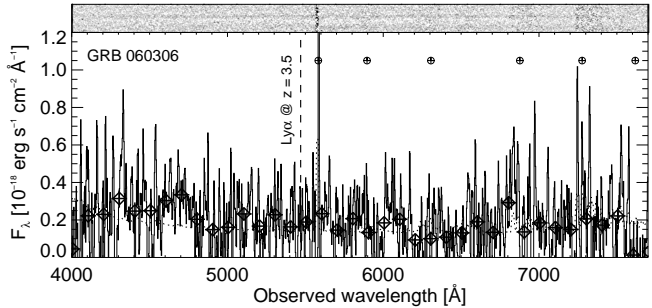}
\caption{One- and two-dimensional spectra (300V) of the GRB\,060306 
host. Telluric features and skyline residuals are marked with $\oplus$, 
whereas the error spectrum is plotted as a dotted line. The diamonds show 
the rebinned (100\,\AA) spectrum. The vertical dashed line indicates the
location of Ly$\alpha$ if $z = 3.5$ \citep{ruben2012}.}
\label{060306.fig}
\end{figure}

\subsection{GRB\,060306 (No OA/NIRA)}

Neither emission lines nor a continuum were detected in the 600z spectrum 
and hence no redshift information could be obtained. However, there is a
faint detection of the continuum in the 300V spectrum in the approximate
wavelength range 4300--6800\,\AA\ (Fig.~\ref{060306.fig}) indicating $0.8 
\lesssim z \lesssim 2.5$. 
\par
In particular, as apparent from the rebinned spectrum (diamonds in 
Fig.~\ref{060306.fig}), we do not detect any flux break corresponding to 
Ly$\alpha$ at $z=3.5$. This redshift was based on the detection of a single 
emission line interpreted as [\ion{O}{2}] \citep{ruben2012}. We note that
this lack of a break is different from what we detect from $z > 2$ hosts
presented in this paper (GRBs 050915A, 051001, and 070129). We tentatively 
suggest that the redshift must be either $z = 2.35$ or $z = 1.55$ if the 
emission line is interpreted as [\ion{O}{3}] or H$\alpha$, respectively.

\subsection{GRB\,060719 (NIRA)}

\begin{figure}
\epsscale{1.17}
\plotone{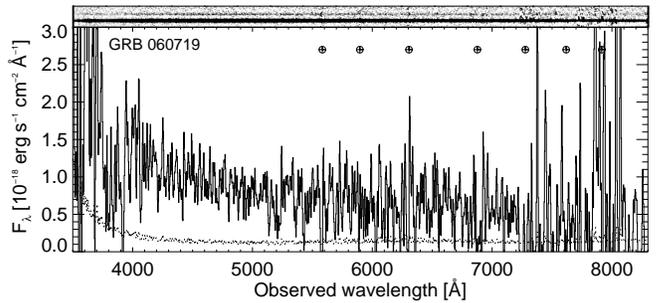}
\caption{One- and two-dimensional spectra (300V) of the GRB\,060719 
host. In the upper panel, it is the fainter upper trace. Telluric features 
and skyline residuals are marked with $\oplus$, whereas the error spectrum 
is plotted as a dotted line.}
\label{060719.fig}
\end{figure}

The continuum is very faint in 600z but slightly brighter in 300V
(Fig.~\ref{060719.fig}). Based on the continuum in the 3600--7200\,\AA\ 
region we infer $0.9 \lesssim z \lesssim 2.0$. Indeed, recent VLT/X-shooter 
observations have detected emission lines (strong H$\alpha$ and an indication 
of [\ion{O}{2}] $\lambda 3727$) outside of our grism wavelength range, 
consistent with the FORS redshift limit \citep{thomas}.  

\subsection{GRB\,060805A (No OA/NIRA)}

There are two host galaxy candidates within the XRT error circle, a bright
object (A) at the southwestern edge and a second, fainter source (B) slightly
northeast of the center \citep{perley,danieleLP}. Our slit covered both objects.
No line features are observed over the spectral range for object B. There
is, however, a faint continuum detectable by block averaging the spectrum 
along the dispersion axis. It is detected down to approximately 4200\,\AA\ 
which corresponds to $z \lesssim 2.5$.
\par
\begin{figure}
\epsscale{1.17}
\plotone{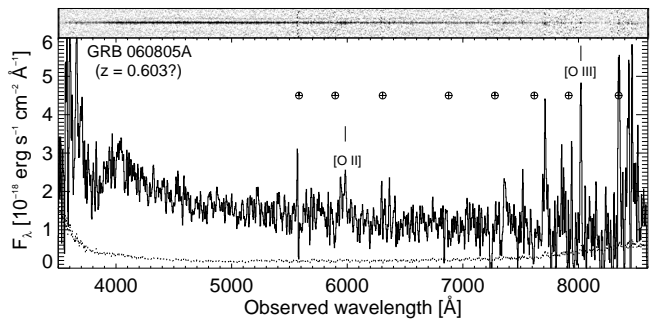}
\caption{One- and two-dimensional spectra (300V) of the brighter galaxy
(object A) within the XRT error circle of GRB\,060805A. The fainter galaxy
(object B) is barely seen in the two-dimensional spectrum just above object A.
Line features are marked with vertical lines, whereas telluric features
and skyline residuals are marked with $\oplus$. The error spectrum is
plotted as a dotted line.}
\label{060805A.fig}
\end{figure}

The continuum of object A is detected down to at least 3600\,\AA\ which 
implies $z \lesssim 2.0$ (Fig.~\ref{060805A.fig}), consistent with the 
findings of \citet{perley}. The spectrum also displays [\ion{O}{2}] 
$\lambda 3727$ and [\ion{O}{3}] $\lambda 5007$ at a common redshift of 
$z = 0.603$. The probability of chance projection for objects A and B
is 1.6\% and 6.6\%, respectively, calculated following the prescription in 
\citet{josh}. Hence, the host identification is ambiguous, and we cannot 
claim to have secured the redshift of this burst.

\begin{figure}
\epsscale{1.17}
\plotone{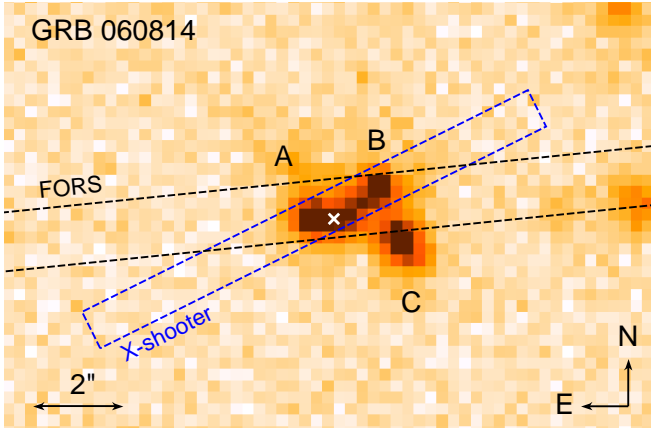}
\caption{$R$-band image (seeing: 0\farcs65) of the GRB\,060814 field. The 
location of the NIR afterglow is marked with a cross and is consistent with 
blob A. The orientation and extent of the FORS slit is also shown; it mostly 
cover blobs A and B, and partly blob C. The X-shooter slit only covers blobs 
A and B.}
\label{060814.fig}
\end{figure}

\subsection{GRB\,060814 (NIRA)}

The host system of GRB\,060814 seems to consist of three different blobs
\citep{dm814} which we mark as A, B, and C (Fig.~\ref{060814.fig}). The 
location of the NIRA \citep{andrew} is consistent with blob A; we 
used the data from \citet{andrew} along with our images to precisely position 
the afterglow on the host complex. \citet{ct} reported a redshift of 
$z = 0.84$ although it is not clear which part of the complex was covered by 
their slit.
\par
\begin{figure}
\epsscale{1.17}
\plotone{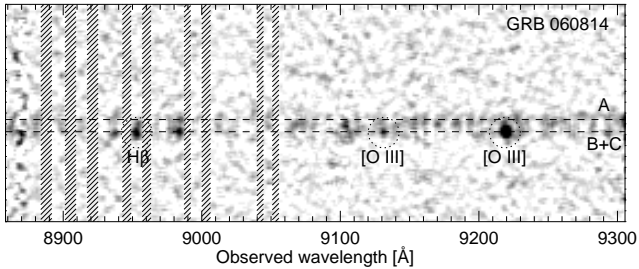}
\caption{Two-dimensional spectrum (600z) of the GRB\,060814 host complex. 
The dashed upper (lower) line indicates the centroid of blob A (B+C) which is 
shown in Fig.~\ref{060814.fig}. Hashed regions mark areas strongly affected by 
skylines.}
\label{060814_600z.fig}
\end{figure}

Our spectra clearly show emission lines emanating from the combined
region of blobs B and C (Fig.~\ref{060814_600z.fig}). In the 600z spectrum 
(seeing: 1\farcs3), we detect H$\beta$ and [\ion{O}{3}] $\lambda \lambda 4959, 
5007$ at a common redshift of $z = 0.841$. There is an indication of 
[\ion{O}{2}] $\lambda 3727$ in the 300V spectrum but residuals from strong sky 
lines make the identification ambiguous with our resolution.
\par
However, there are no indications of any emission lines at the location of 
blob A. A clear detection of the continuum in the 300V spectrum in the 
approximate range 3600--7200\,\AA\ suggests that $0.9 \lesssim z \lesssim 2.0$.
In any case, the conclusion is that a host redshift of $z = 0.84$ is not 
supported by our data. Indeed, recent VLT/X-shooter observations led to the 
detection of [\ion{O}{2}] $\lambda 3727$, [\ion{O}{3}] $\lambda \lambda 
4959,5007$, and H$\alpha$ in emission at $z = 1.92$ \citep{thomas}, clearly 
visible in blobs A and B (the slit did not cover blob C as shown in 
Fig.~\ref{060814.fig}).\footnote{Our analysis is consistent with what is
reported by \citet{ruben2012} who used the same dataset.} At this redshift, 
the most prominent emission lines are outside of our FORS grisms wavelength 
range. The necessary consequence of these observations taken together is that 
the host complex consists of blobs A and B at $z = 1.92$, while blob C is a 
chance alignment at $z = 0.84$.

\subsection{GRB\,060908 (OA)}

\begin{figure}
\epsscale{0.45}
\plotone{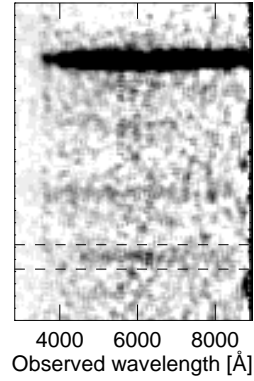}
\caption{Two-dimensional spectrum (300V) of the GRB\,060923A host. It has been
block averaged along the dispersion axis to make the host continuum visible.
It is the bottom trace between the two horizontal dashed lines.}
\label{060923A.fig}
\end{figure}

A Ly$\alpha$ emission line is detected from the host galaxy at a redshift of 
$z = 1.884$ \citep[part of the TOUGH Ly$\alpha$ campaign:][]{boLP}. 
\citet{johan_sample} utilized these findings and confirmed the redshift by 
detecting the \ion{C}{4} $\lambda \lambda 1548, 1550$ doublet in the OA 
spectrum, ruling out the tentative redshift of $z = 2.43$ reported by 
\citet{rol}.

\subsection{GRB\,060923A (NIRA)}

The extremely red afterglow of GRB 060923A was discussed in \citet{nial}. 
Their later-time optical imaging and spectroscopy revealed a faint galaxy 
coincident with the GRB position which implied a moderate redshift and 
therefore that dust is the likely cause of the very red afterglow color.
\par
Our 300V spectrum reveals a very faint continuum with a possible weak 
emission line candidate at around 6330\,\AA\ (3$\sigma$ detection). In order 
to verify this we obtained a higher resolution 1200R spectrum. There is again
a weak detection of the continuum but no sign of an emission line at the 
aforementioned wavelength. In Fig.~\ref{060923A.fig} the continuum
is detected down to approximately 4600\,\AA\ which corresponds to 
$z \lesssim 2.8$ \citep{nial}.

\subsection{GRB\,060923C (NIRA)}

\label{060923C.sec}

\begin{figure}
\epsscale{2.34}
\plottwo{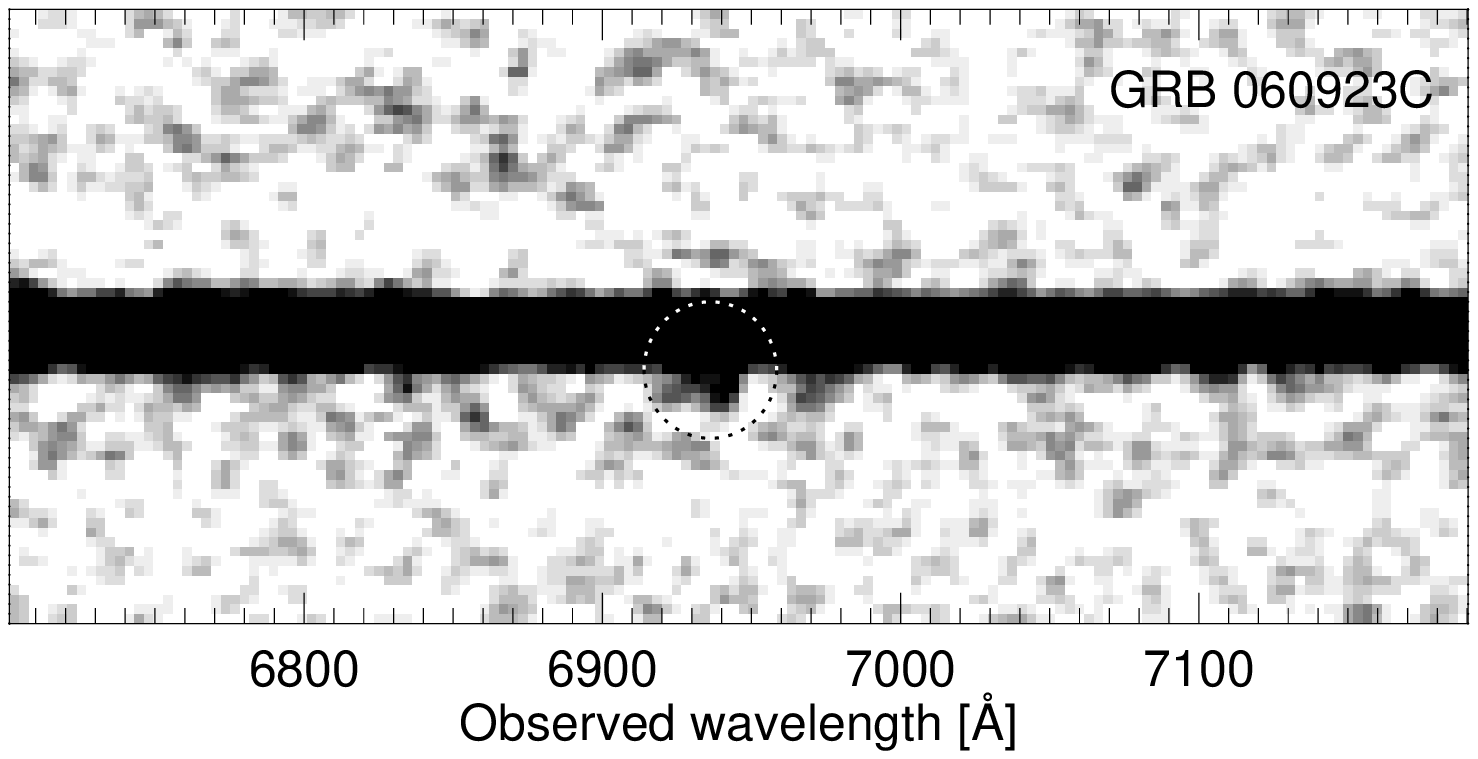}{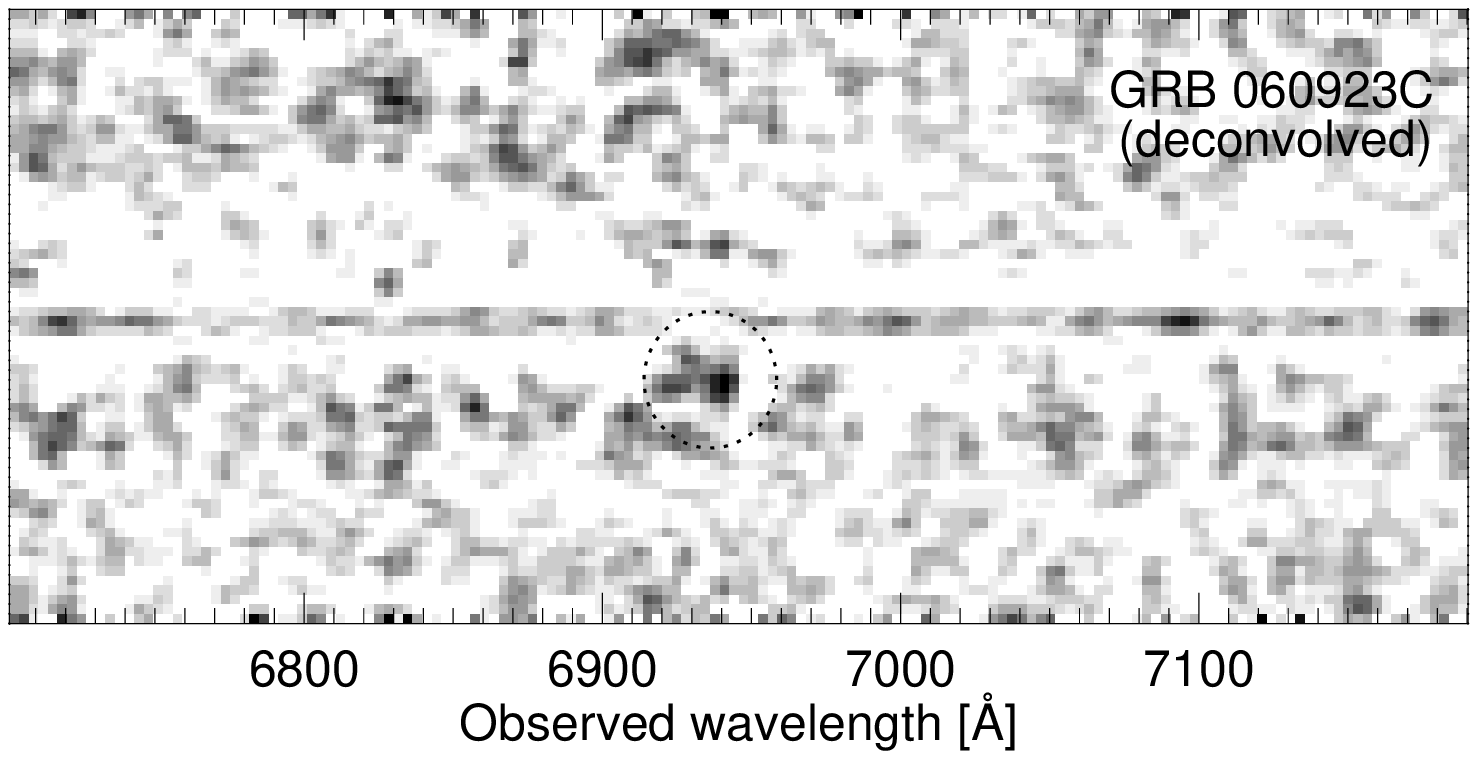}
\caption{Original (upper panel) and deconvolved (lower panel) 
two-dimensional spectrum (300V) of the GRB\,060923C host. The emission 
line candidate is indicated with a circle.}
\label{060923C.fig}
\end{figure}

The GRB\,060923C NIRA was located only around 1\farcs5 away
from an $R = 20.9$\,mag bright star. Detecting the host ($R = 25.5$\,mag) 
spectrum, even under good seeing conditions, is therefore quite a challenge. 
As can be seen in the upper panel of Fig.~\ref{060923C.fig}, the stellar 
spectrum overwhelms any signs of the host continuum. However, there is an 
indication of an emission line at the expected location of the host trace at 
approximately 6942\,\AA. In order to get a better view of the situation we 
subtracted the star continuum using a two-dimensional deconvolution technique.
\par
The deconvolution method applied to the spectrum is presented in detail 
in \citet{courbin}, resulting in a spatially deconvolved spectrum with the
stellar spectrum separated from the spectrum of the host galaxy. This
spatial separation of the pointlike component from the diffuse one relies
on an accurate building of the point-spread function and on the hypothesis 
that the host contains no significant structure narrower than the fixed and 
finite resolution of the deconvolved spectrum. Following \citet{letawe08},
we have chosen to subtract the stellar spectrum, as derived from the 
deconvolution process, from the original spectrum. This avoids smoothing of 
the diffuse component mandatory for a proper separation between the star and 
host. Thus, we obtain a two-dimensional host galaxy spectrum with a spatial 
resolution constant along the slit and equal to that of the original data
\par
The results are shown in the lower panel of Fig.~\ref{060923C.fig}. The 
emission line candidate is still visible. If real, the identification could 
only be with Ly$\alpha$ or [\ion{O}{2}] $\lambda 3727$ in which case the 
redshift would be $z = 4.71$ or $z = 0.86$, respectively. We have considered 
other potential line identifications such as H$\beta$, [\ion{O}{3}] $\lambda 
5007$, or H$\alpha$ all of which are rejected on account that we would expect 
to see additional and roughly similarly bright lines in the wavelength range 
covered. We note that the \citet{grupe} method implies that $z \lesssim 3.5$
so that $z = 4.71$ is excluded.

\subsection{GRB\,061021 (OA)}

\begin{figure}
\epsscale{1.17}
\plotone{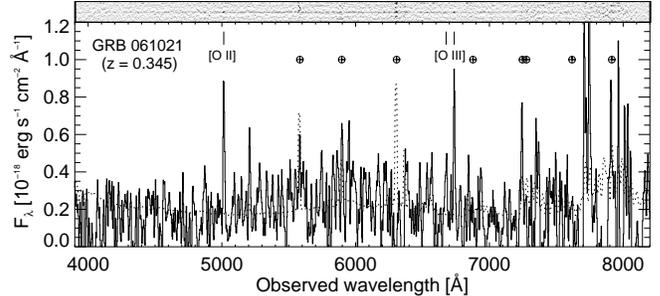}
\caption{One- and two-dimensional spectra (300V) of the GRB\,061021 host. 
Emission lines are marked with vertical lines, whereas telluric features and 
skyline residuals are marked with $\oplus$. The error spectrum is plotted as 
a dotted line.}
\label{061021.fig}
\end{figure}

Our 300V host spectrum shows [\ion{O}{2}] $\lambda 3727$ and [\ion{O}{3}] 
$\lambda \lambda 4959, 5007$ at a common redshift of $z = 0.345$ 
(Fig.~\ref{061021.fig}). Using this redshift information, \citet{johan_sample} 
identified the \ion{Mg}{2} $\lambda \lambda 2796,2803$ doublet in the very 
blue end of the afterglow spectrum.

\subsection{GRB\,070103 (No OA/NIRA)}

\begin{figure}
\epsscale{1.17}
\plotone{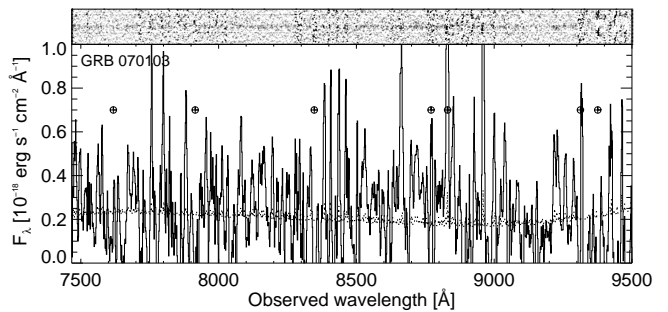}
\caption{One- and two-dimensional spectra (600z) of the GRB\,070103 
host. Telluric features and skyline residuals are marked with $\oplus$, 
whereas the error spectrum is plotted as a dotted line.}
\label{070103.fig}
\end{figure}

The host continuum detection down to 7500\,\AA\ in our 600z spectrum 
only allows us to set a modest limit of $z \lesssim 5.2$ 
(Fig.~\ref{070103.fig}). This is less constraining than using the 
\citet{grupe} method which implies that $z \lesssim 3.5$ \citep{johan_sample}. 
There are no convincing emission lines present. However, recent VLT/X-shooter 
observations have detected [\ion{O}{3}] $\lambda \lambda 4959,5007$ in 
emission outside of our grism wavelength range, consistent with the 
aforementioned redshift limits \citep{thomas}.  

\subsection{GRB\,070129 (OA)}

\begin{figure}
\epsscale{1.17}
\plotone{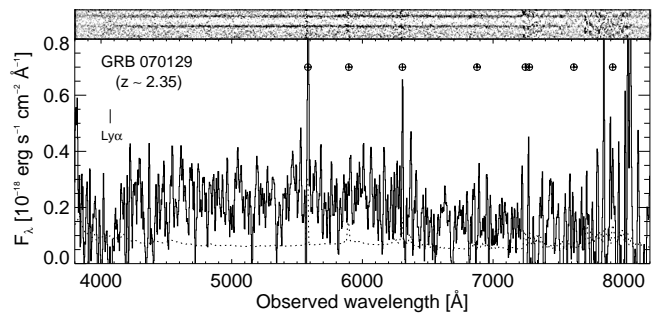}
\caption{One- and two-dimensional spectra (300V) of the GRB\,070129
host (lower trace). The Ly$\alpha$ absorption feature is marked with a 
vertical line, whereas telluric features and skyline residuals are marked with 
$\oplus$. The error spectrum is plotted as a dotted line.}
\label{070129.fig}
\end{figure}

Neither emission lines nor a continuum were detected in the 600z spectrum 
and hence no redshift information could be obtained. However, there is a
clear detection of the continuum in the 300V spectrum between approximately
4100 and 7200\,\AA, suggesting that $0.9 \lesssim z \lesssim 2.4$. Thus, we
can immediately rule GRB\,070129 out as a high-redshift candidate as suggested 
by \citet{salvaZ}.
\par
In the upper panel of Fig.~\ref{070129.fig}, the continuum of another
similarly bright object on the slit is visible. It clearly contains flux 
down to around 3900\,\AA. Therefore, a spectral break is present in the 
GRB\,070129 host continuum, which we interpret as the Ly$\alpha$ break. More 
specifically, the 1D spectrum shows a flux drop around 4070\,\AA\ 
corresponding to $z \approx 2.35$. Recent observations by VLT/X-shooter 
confirm this interpretation where we detect [\ion{O}{2}] $\lambda 3727$, 
[\ion{O}{3}] $\lambda \lambda 4959, 5007$, and H$\alpha$ in emission at a
similar redshift \citep{thomas}.

\subsection{GRB\,070306 (NIRA)}

The highly extinguished afterglow of GRB 070306 and the properties of its 
host galaxy were discussed in \citet{andreas}. The OA low-resolution spectrum 
displayed a single emission line which was tentatively interpreted as 
[\ion{O}{2}] $\lambda 3727$. In an attempt to verify this interpretation by 
resolving the doublet we observed the host galaxy with the higher-resolution 
1028z grism as part of TOUGH. As shown in Figure~4 in \citet{andreas} the 
doublet was successfully resolved and a redshift of $z = 1.496$ confirmed.

\subsection{GRB\,070328 (No OA/NIRA)}

A very faint host continuum detection down to 7500\,\AA\ in our 600z spectrum 
only allows us to set a modest limit of $z \lesssim 5.2$. This is less 
constraining than using the \citet{grupe} method which implies that 
$z \lesssim 3.5$ \citep{johan_sample}. There are no convincing emission lines 
present.


\begin{deluxetable*}{@{}lcrlllcccr@{}}
\tablecaption{New Redshifts and Redshift Limits Determined in the Host
Galaxy Sample \label{z.tab}}
\tablewidth{0pt}
\tablehead{
GRB &
OA or NIRA? &
Redshift &
$R_{\textrm{host}}$ &
Instrument &
Grism$+$Filter & 
Slit Width &
Seeing     &
Spectral Res. &
Exp. Time \\
 &
 &
 &
(mag) &
 &
 &
(\arcsec) &
(\arcsec) &
(\AA)     &
(s) \hspace{3mm}
}

\startdata

\multirow{2}{*}{050714B} & \multirow{2}{*}{No} & 
\multirow{2}{*}{$z \lesssim 3.5$} & \multirow{2}{*}{$25.5$}

& oFORS1 & 300V          & $1.3$ & 0.9 & 11.9 \hspace{0.4mm} & $4 \times 1220$ \\
                         
                         &       &     &
& nFORS1 & 300V          & $1.3$ & 0.7 &  7.3 & $4 \times 1220$ \\

\multirow{2}{*}{050822}  & \multirow{2}{*}{No} & \multirow{2}{*}{$1.434$} &
\multirow{2}{*}{$24.2$}
& oFORS1 & 300V+GG375    & $1.3$ & 0.8 & 10.6 \hspace{0.4mm} & $2 \times 1300$ \\

                         &       &     &
& FORS2  & 600z+OG590    & $1.0$ & 1.9 & 3.2 & $2 \times 1275$ \\

050915A                  & NIRA & $z \approx 2.54$ & $24.6$
& oFORS1 & 300V          & $1.3$ & 1.1 & 14.5 \hspace{0.4mm} & $4 \times 1295$ \\

\multirow{2}{*}{051001}  & \multirow{2}{*}{No} & 
\multirow{2}{*}{$z \approx 2.43$} & \multirow{2}{*}{$24.4$}

& nFORS1 & 300V          & $1.3$ & 0.8 & 8.4 & $4 \times 1360$ \\
                         
                         &       &     &
& FORS2  & 600z+OG590    & $1.0$ & 0.7 & 2.3 & $2 \times 1350$ \\

051006                   & No    & $1.059$   & $23.0$
& FORS2  & 600RI+GG435   & $1.0$ & 1.3 & 3.3 & $2 \times 430$  \\

051117B                  & No    & $0.481$   & $21.1$
& oFORS1 & 300V          & $1.3$ & 1.3 & 17.2 \hspace{0.4mm} & $1 \times 1000$ \\

\multirow{2}{*}{060306}  & \multirow{2}{*}{No} & 
\multirow{2}{*}{$0.8 \lesssim z \lesssim 2.5$} & \multirow{2}{*}{$24.1$}
 
& oFORS1 & 300V+GG375    & $1.3$ & 1.3 & 17.2 \hspace{0.4mm} & $2 \times 1300$ \\
                         
                         &       &     &
& FORS2  & 600z+OG590    & $1.0$ & 1.1 & 3.2 & $2 \times 1275$ \\

\multirow{2}{*}{060719}  & \multirow{2}{*}{NIRA} & 
\multirow{2}{*}{$0.9 \lesssim z \lesssim 2.0$}   & \multirow{2}{*}{$24.6$}

& nFORS1 & 300V          & $1.3$ & 0.9 & 9.4 & $4 \times 1360$ \\
                         
                         &       &     &
& FORS2  & 600z+OG590    & $1.0$ & 1.0 & 3.2 & $4 \times 1350$ \\

\multirow{2}{*}{060805A} & \multirow{2}{*}{No} & 
\multirow{2}{*}{$z \lesssim 2.5$} 

& 23.5 (A) & nFORS1 & 300V & $1.3$ & 1.3 & 13.6 \hspace{0.4mm} & $4 \times 1295$ \\
                         
                         &       &       &
25.1 (B) & FORS2  & 600z+OG590   & $1.0$ & 1.4 & 3.2 & $1 \times 1500$ \\

\multirow{2}{*}{060814}  & \multirow{2}{*}{NIRA} & 
\multirow{2}{*}{$1.92$}  & \multirow{2}{*}{$22.9$}

& nFORS1 & 300V          & $1.3$ & 0.8 & 8.4 & $2 \times 1295$ \\
                         
                         &       &     &
& FORS2  & 600z+OG590    & $1.0$ & 1.3 & 3.2 & $2 \times 1350$ \\

060908                   & OA   & $1.884$ & $25.5$
& nFORS1 & 600B          & $1.3$ & 0.8 & 6.5 & $4 \times 1345$ \\

\multirow{2}{*}{060923A} & \multirow{2}{*}{NIRA} & 
\multirow{2}{*}{$z \lesssim 2.8$} & \multirow{2}{*}{$26.1$}

& nFORS1 & 300V          & $1.3$ & 1.0 & 10.5 \hspace{0.4mm} & $2 \times 1280$ \\
                         
                         &       &     &
& FORS2  & 1200R+GG435   & $1.3$ & 1.0 & 1.5 & $2 \times 1350$ \\

060923C                  & NIRA  & $z \lesssim 3.5$ & $25.5$
& nFORS1 & 300V          & $1.0$ & 0.8 & 8.4 & $4 \times 1360$ \\

061021                   & OA    & $0.345$   & $24.4$
& nFORS1 & 300V          & $1.3$ & 0.7 & 7.3 & $2 \times 1280$ \\

070103                   & No    & $z \lesssim 3.5$ & $24.2$
& FORS2  & 600z+OG590    & $1.0$ & 1.0 & 3.2 & $4 \times 900$  \\

\multirow{2}{*}{070129}  & \multirow{2}{*}{OA} & 
\multirow{2}{*}{$z \approx 2.35$} & \multirow{2}{*}{$24.4$}

& nFORS1 & 300V          & $1.3$ & 1.4 & 13.6 \hspace{0.4mm} & $4 \times 1360$ \\
                         
                         &       &     &
& FORS2  & 600z+OG590    & $1.0$ & 1.4 & 3.2 & $2 \times 1275$ \\

070306                   & NIRA  & $1.496$   & $22.9$
& FORS2  & 1028z+OG590   & $1.0$ & 0.7 & 2.5 & $2 \times 1300$  \\

070328                   & No    & $z \lesssim 3.5$ & $24.4$
& FORS2  & 600z+OG590    & $1.0$ & 1.0 & 3.2 & $5 \times 1330$  \\

070419B                  & OA   & $0.9 \lesssim z \lesssim 2.2$ & $25.2$
& nFORS1 & 300V          & $1.3$ & 0.9 & 9.4 & $6 \times 1360$ \\

070808                   & No    & $z \lesssim 3.5$ & \hspace*{-3.3mm} 
$\gtrsim$$26.7$
& FORS2  & 600RI+GG435   & $1.3$ & 2.2 & 4.3 & $1 \times 715$  \\

\vspace{-6 mm} \\

\enddata

\tablecomments{$R_{\textrm{host}}$ is the $R$-band total magnitude (or 
3$\sigma$ upper limit) of the host galaxy (before correcting for Galactic 
extinction) from \citet{danieleLP}. The GRB\,060805A host redshift limit 
is conservative and refers to the fainter object (B) within the XRT error 
circle. The GRB\,060908 host redshift originates from the Ly$\alpha$ part 
of TOUGH \citep{boLP}. The host redshifts of GRBs 060719, 070103, and 070419B 
are reported in \citet{thomas} and stem from our VLT/X-shooter observations. 
The redshift uncertainty is of the order of $\pm 0.001$ for those
given to three decimal places. Data obtained with FORS1 before 2007 April 6 
are marked oFORS1. Data obtained after the FORS1 blue CCD upgrade are 
labeled nFORS1. The wavelength coverage for each setup is approximately 
3500--8140\,\AA\ (oFORS1/300V/1\farcs3), 
3500--9640\,\AA\ (nFORS1/300V/1\farcs0), 
3500--8880\,\AA\ (nFORS1/300V/1\farcs3),
7470--10\,700\,\AA\ (FORS2/600z/1\farcs0), 
5300--8630\,\AA\ (FORS2/600RI/1\farcs0),
4950--8250\,\AA\ (FORS2/600RI/1\farcs3), 
7700--9500\,\AA\ (FORS2/1028z/1\farcs0), 
and 5720--7200\,\AA\ (FORS2/1200R/1\farcs3).}
\end{deluxetable*}

\subsection{GRB\,070419B (OA)}

\begin{figure}
\epsscale{1.17}
\plotone{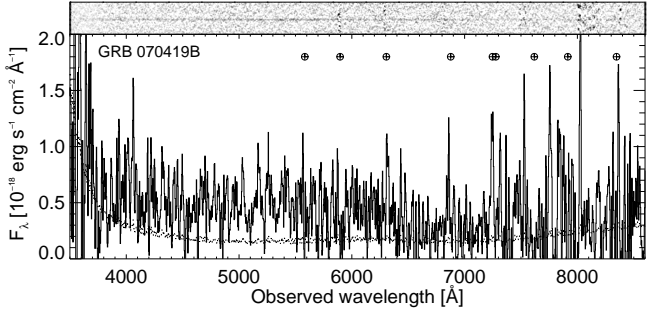}
\caption{One- and two-dimensional spectra (300V) of the GRB\,070419B 
host. Telluric features and skyline residuals are marked with $\oplus$, 
whereas the error spectrum is plotted as a dotted line.}
\label{070419B.fig}
\end{figure}

The continuum is detected in the 300V spectrum but without any significant
emission lines visible (Fig.~\ref{070419B.fig}). There is an unambiguous
detection of the continuum between approximately 3900 and 7200\,\AA\ indicating
that $0.9 \lesssim z \lesssim 2.2$. Indeed, recent VLT/X-shooter observations 
have detected [\ion{O}{3}] $\lambda \lambda 4959,5007$ and H$\alpha$ in 
emission outside of our grism wavelength range, consistent with the FORS 
redshift limit \citep{thomas}.  

\subsection{GRB\,070808 (No OA/NIRA)}

\begin{figure}
\epsscale{1.17}
\plotone{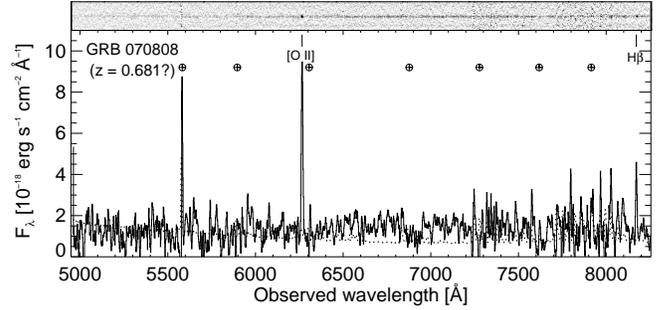}
\caption{One- and two-dimensional spectra (600RI) of object A located 
close to the NE border of the GRB\,070808 XRT error circle. Emission lines 
are marked with vertical lines, whereas telluric features and skyline 
residuals are marked with $\oplus$. The error spectrum is plotted as a dotted 
line.}
\label{070808.fig}
\end{figure}

There is a faint ($R = 26.7$) object consistent with the X-ray position, as 
well as two brighter ones just outside the X-ray error circle. We obtained a 
600RI spectrum of the brightest source \cite[object A:][]{danieleLP}. It 
clearly shows two emission lines which we identify as the [\ion{O}{2}] 
$\lambda 3727$ doublet and H$\beta$ at a common redshift of $z = 0.681$ 
(Fig.~\ref{070808.fig}). The probability of chance projection for this object 
is around 4\%--5\%, calculated following the prescription in \citet{josh}, 
moreover it definitely lies outside the UVOT-enhanced XRT error circle. The 
fainter object is consistent with the XRT error circle, but it also has a 
large chance superposition probability (25\%). Hence, we cannot claim to have 
secured the redshift of this burst. However, we conclude that $z \lesssim 3.5$ 
based on the excess column density detected in the X-ray spectrum.

\begin{deluxetable}{@{}llrc@{}}
\tablecaption{Line Identifications, Observed EWs, and Fluxes \label{lines.tab}}
\tablewidth{0pt}
\tablehead{
GRB &
Feature &
EW$_{\textrm{obs}}$ \hspace{1mm} &
Flux ($10^{-17}$ \\
 &
 &
(\AA) \hspace{3.5mm} &
erg\,s$^{-1}$\,cm$^{-2}$)
}

\startdata

050822                   & [\ion{O}{2}]  $\lambda 3727$ & 
$< -23$ \hspace{8.7mm} & $2.2$ \\
\hline
\multirow{2}{*}{051006}  & [\ion{O}{2}]  $\lambda 3727$ & 
$-95.7 \pm 4.0$ & $7.5$ \\
                         & [\ion{Ne}{3}] $\lambda 3869$ & 
$-28.4 \pm 2.9$ & $2.0$ \\
\hline
\multirow{6}{*}{051117B} & [\ion{O}{2}]  $\lambda 3727$ & 
$-12.5 \pm 1.5$ & $6.6$ \\
                         & H10                          & 
$  5.0 \pm 1.1$ & ---   \\
                         & H9                           & 
$  6.8 \pm 1.1$ & ---   \\
                         & H8                           & 
$  6.0 \pm 1.4$ & ---   \\
                         & \ion{Ca}{2} K                & 
$  5.2 \pm 1.1$ & ---   \\
                         & \ion{Ca}{2} H and H$\varepsilon$ & 
--- \hspace{3mm} & ---   \\
\hline
\multirow{2}{*}{060805A} & [\ion{O}{2}]  $\lambda 3727$ & 
$-20.0 \pm 4.0$          & 2.6 \\
                         & [\ion{O}{3}]  $\lambda 5007$ &       
$-67.0 \pm 9.0$          & 7.6  \\
\hline
\multirow{3}{*}{061021}  & [\ion{O}{2}]  $\lambda 3727$ & 
$-49.1 \pm 9.9$ & $1.5$ \\
                         & [\ion{O}{3}]  $\lambda 4959$ & 
$-24.0 \pm 9.0$ & $0.8$ \\
                         & [\ion{O}{3}]  $\lambda 5007$ & 
$-51.0 \pm 9.5$ & $1.3$ \\
\hline
\multirow{2}{*}{070306}  & [\ion{O}{2}]  $\lambda 3726$ & 
--- \hspace{3mm}       & \multirow{2}{*}{17.0} \hspace{0.5mm} \\
                         
                         & [\ion{O}{2}]  $\lambda 3729$ &       
--- \hspace{3mm}       &                              \\
%
\hline

\multirow{2}{*}{070808}  & [\ion{O}{2}]  $\lambda 3727$ & 
$-65.3 \pm 5.1$ & $10.9$ 
\hspace{0.5mm} \\
                         & H$\beta$                     & 
$-17.5 \pm 3.3$ & $4.5$ \\

\vspace{-6 mm} \\

\enddata

\tablecomments{Flux measurement errors are of the order of 20\%. When the
EW is not given, skylines prohibited a reliable estimate. The fluxes 
reported for GRB\,060805A are those from object A. The fluxes reported
for GRB\,070808 are those from object A. The flux in the emission line 
candidate in GRB\,060923C is not reported due to the uncertainties introduced 
by the deconvolution.}

\end{deluxetable}

\begin{figure*}
\epsscale{1.1}
\plotone{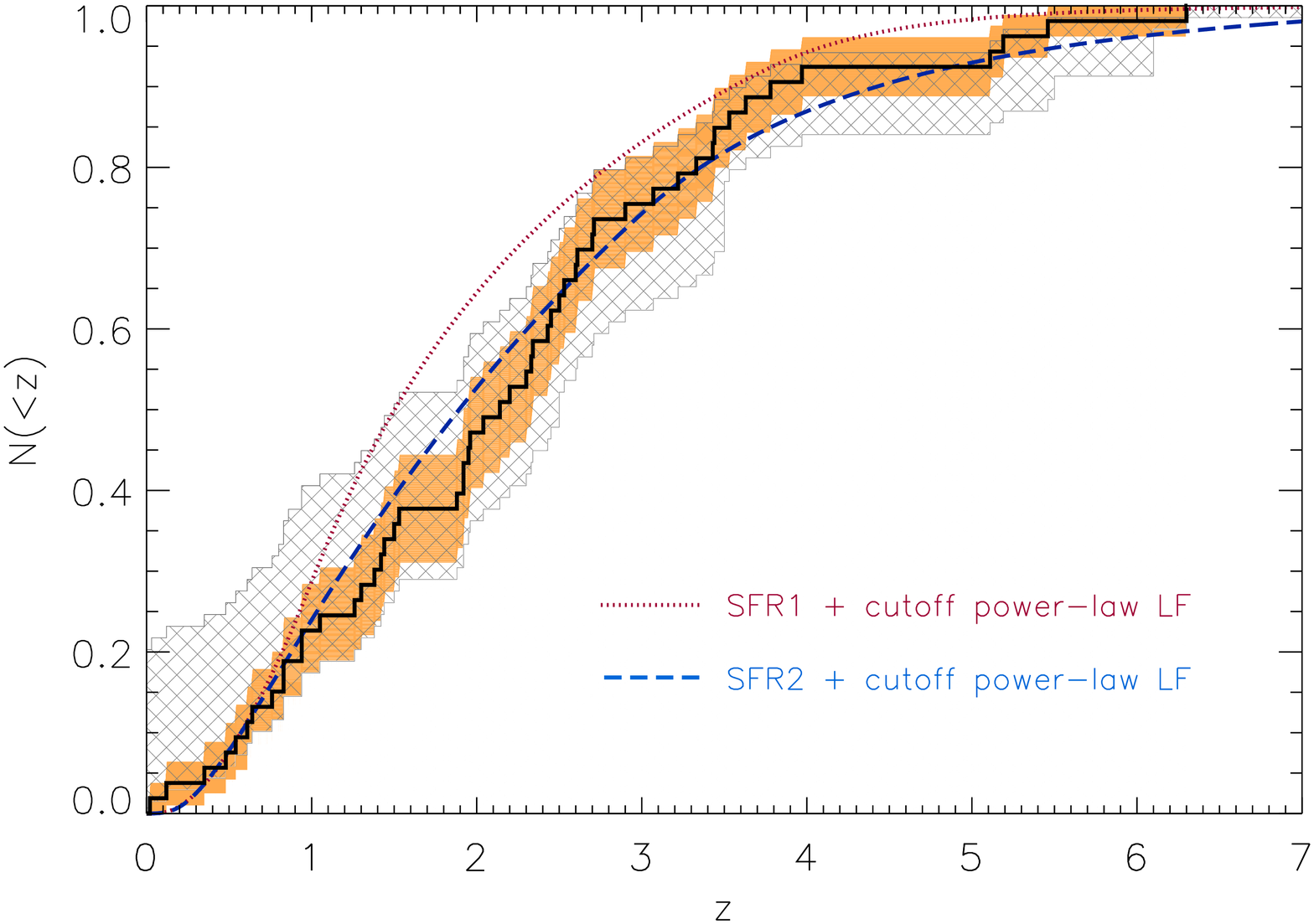}
\caption{{\bf Thick solid curve}: the cumulative fraction of GRBs as a 
    function of redshift for the 53 \emph{Swift} bursts in the TOUGH sample 
    with a measured redshift ($\langle z \rangle = 2.23$). 
    {\bf Hatched region}: this is a conservative error region showing the 
    systematic error on the thick solid curve. 
    {\bf Shaded region}: statistical region showing the 1$\sigma$ sampling 
    error band around the thick solid curve. 
    {\bf Dotted curve}: the expected redshift distribution for \emph{Swift} 
    observable long GRBs using the \emph{SFR1} history parameterization (see 
    the main text). 
    {\bf Dashed curve}: the same redshift distribution for the \emph{SFR2} 
    history parameterization (see the main text).}
\label{zdist.fig}
\end{figure*}

\subsection{Summary}

\label{sum.sec}

We report the redshifts of 10 GRB host galaxies whose average redshift is 
$\langle z \rangle = 1.59$, significantly lower than the overall \textit{Swift} 
GRB mean redshift. The low value is most likely the result of targeting the 
brightest galaxies in the sample ($R \lesssim 25$\,mag). GRB\,061021 is among 
the closest ``classical'' \emph{Swift} long-duration burst detected so far. 
Furthermore, we have estimated redshift limits for an additional five hosts 
and inferred that three burst redshifts reported in the literature are 
erroneous (GRBs 060306, 060814, and 060908). The results are listed in 
Table~\ref{z.tab} along with the observational details. We have also estimated 
the emission line flux and the equivalent width (EW) of most line features 
detected and detailed them in Table~\ref{lines.tab}. 

\section{THE GRB REDSHIFT DISTRIBUTION}

\label{dist.sec}

\subsection{Redshift Measurements and Constraints}

Figure~\ref{zdist.fig} shows the cumulative redshift distribution of the 
53 TOUGH bursts with a measured redshift, including four photometric redshifts 
and the new VLT/X-shooter redshifts presented in \cite{thomas} \citep[see 
Table~3 in][]{jensLP}. Also plotted is a conservative systematic error band 
(hatched region) containing information for all the 69 TOUGH bursts. The 
shaded region represents the likely statistical (1$\sigma$ standard error of 
the sample) uncertainty of the measured redshift distribution under the 
assumption that it is a true random sample of the overall population.
\par
The hatched region incorporates the information presented in Table~\ref{z.tab},
including the redshift limits. In addition, the upper boundary is produced by 
simply placing GRBs with only upper limits to their redshifts at $z = 0$. The 
lower boundary is generated in the following way. GRBs with a detected OA 
and/or host are placed at the maximum redshift they can have, given their 
bluest photometric detection reported in the literature or in \cite{jensLP}. 
Following the general decline of the Ly$\alpha$ forest density with decreasing 
redshift \citep[e.g.][]{son}, we set two different limits depending on the 
wavelength region probed by the available observations. At low redshift, the 
opacity of the Ly$\alpha$ forest is small, hence, for detections in blue and 
optical filters ($V$ band and blueward), the maximum redshift is derived by 
comparing the reddest wavelength in the filter response curve to the Lyman 
limit. At higher redshift, only negligible flux survives blueward of the 
Ly$\alpha$ wavelength, therefore, for redder filters, the comparison is to 
Ly$\alpha$ \citep[e.g.][]{dall}. If a more constraining limit is not available,
we can set their maximum redshift to $z = 3.5$ if their X-ray spectra fulfill 
the criterion described in \S \ref{z.sec}. For only a single burst, 
GRB\,061004, are there no redshift constraints available. It is arbitrarily 
placed at a maximum redshift of $z = 10$ in Fig.~\ref{zdist.fig}.
\subsection{Modelling}
Various authors have attempted to predict or model the GRB redshift 
distribution and hence compare it to observations \citep[e.g.][]{javier,priya,
palli28,le,salvaKING,guetta07,np,kistler,dong,maria,robellis}, but of necessity 
have worked with samples with considerably greater incompleteness and optical 
bias than TOUGH now provides. During the review process of this work, 
\citet{ruben2012} presented the redshift distribution of a complete sample of 
\emph{Swift} GRBs, which was inspired by criteria similar to those applied to
TOUGH. Their sample is slightly smaller and, by construction, limited to 
gamma-ray bright events. The TOUGH sample extends to fainter luminosities (a 
factor of around six) and as such is more suited to test the faint (and 
potentially high-redshift) end of the GRB distribution. Indeed, their mean 
(median) redshift is 1.84 (1.64), significantly lower than for the whole TOUGH 
sample.
\par
Given the association of long GRBs with the deaths of massive stars 
\citep[e.g.][]{jens,WB}, it is commonly assumed that the GRB rate density 
follows the star-formation rate (SFR) density history \citep[e.g.][]{pacz,
ralph,porc,palli05,johan}, with a possible low-metallicity enhancement 
\citep[e.g.][]{langer,li,butler} and/or evolution with redshift in intrinsic 
rate or LF \citep[e.g.][]{coward05,daigne,guetta07,kistler,salva}. For the 
purposes of this observational paper, we will only present illustrative model 
fits.
\par
Currently, we have little guidance from theory as to plausible functional forms
for the GRB LF, and various ones have been considered in the above studies.
Two of the most commonly adopted forms are either a broken power law 
\citep[e.g.][]{guetta05,guetta07,butler,wand}, or a single power law with an 
exponential cutoff at low luminosities \citep[e.g.][]{priya,maria,cao}. Both 
forms generally underpredict the rate of low-luminosity GRBs (as exemplified 
by GRB\,980425), suggesting that such bursts may form a separate population 
\citep[][]{cobb,sod,liang,coward08,zitouni,foley,virgili}. In our homogeneous 
and unbiased TOUGH sample, only one event (GRB\,060218) is generally regarded 
as a member of this faint population, though we note this represents a rate of 
nearby bursts consistent with that predicted from the BATSE sample by 
\citet{bob}. For the sake of clarity and simplicity, we restrict ourselves 
here to the exponentially cutoff single power-law form:

\begin{equation}
\phi(L)\propto (L/L_p)^{-\nu}e^{-L_p/L}.
\end{equation}

\par
We assume that the GRB rate follows the SFR history, and consider two 
different SFR history parameterizations which we label as follows. 
\emph{SFR1} is an update \citep{li} of the SFR history models of \citet{hop}
to include recent data from \citet{bouwens} and \citet{reddy}, combined with a 
low-metallicity modification following the prescription of \citet{langer}. 
\emph{SFR2} is model A from \citet{schmidt09} which represents a SFR history 
which remains constant beyond $z \sim 3$. It may, for example, be considered a 
more extreme low-metallicity correction to the cosmic SFR, or represent a 
correction \citep{kistler09,virgili11} to the high-redshift SFR as estimated 
from flux-limited surveys (by the integration of galaxy LFs thus obtained) due 
to a large amount of hidden star formation in faint, low-mass, and high 
specific SFR galaxies of the type that GRBs tend to be associated with at lower 
redshift \citep[e.g.][]{johan02,lefl,lise,andy}.
\par
Modeling is performed in the standard manner \citep[e.g.][]{guetta07} to 
produce $\log N$-$\log L$ number count distributions for various parameters 
of the LF, which are then fit by $\chi^2$ minimization to the observed 
$\log N$-$\log L$ distribution of all \emph{Swift} bursts with peak photon 
flux $>1{\rm\, cm^{-2}\,s^{-1}}$. We emphasize that the redshift distribution 
is not part of this fitting procedure, but is always purely a result. In 
Fig.~\ref{zdist.fig}, we plot the redshift distributions from our best 
fitting models in comparison to the TOUGH redshift data. The best fit 
model incorporating the \emph{SFR1} parameterization is a good fit to the 
number count distribution with a reduced $\chi^2 = 1.32$ (12 dof) and has 
parameters $L_p = 10^{49.88}$\,erg\,s$^{-1}$ and $\nu = 1.82$. The resultant 
redshift distribution is a good fit to the TOUGH redshift distribution up to 
$z \sim 1.5$, but underpredicts bursts at higher redshift. Overall, a 
Kolmogorov-Smirnov (KS) test yields a $<$0.5\% likelihood of the observed 
and model populations being the same. The \emph{SFR2} parameterization is a 
slightly better fit to the number count distribution with a reduced 
$\chi^2 = 1.30$ (12 dof) and LF parameters $L_p = 10^{50.23}$\,erg\,s$^{-1}$ 
and $\nu = 1.83$. The resultant redshift distribution in this case is an
improved match to the overall observed TOUGH distribution with a KS 
likelihood of 42\%.
\par
At face value, these results seem to imply that GRBs follow a cosmic SFR
history that is significantly enhanced at high redshift compared to estimates
from flux-limited surveys. As previously discussed, given what is known about
GRB hosts, it is entirely feasible that GRBs trace star formation at high 
redshift that would be undetectable by other means. It is of course also 
possible that the simple low-metallicity enhanced SFR parameterization used 
in the \emph{SFR1} model is inadequate, or that the LF could have a more 
complex form and/or evolve with redshift.
\par
An alternative approach to the modeling is to directly fit the observed
joint peak-luminosity and redshift distribution of the sub-sample of TOUGH
GRBs with redshifts. Again we take both \emph{SFR1} and \emph{SFR2}, but in
this case a broken power-law LF. We take a threshold flux for detection by 
the Burst Alert Telescope of $2\times10^{-8}$\,erg\,cm$^{-2}$\,s$^{-1}$, and 
apply a $K$-correction based on the observed spectral parameters for each 
burst, to bring the luminosity to a fixed restframe bandpass (30--300\,keV). 
The results are very similar to the conclusions obtained above; hence we do not 
plot the inferred redshift distribution in Fig.~\ref{zdist.fig}. However, we 
do note that for \emph{SFR2} the maximum-likelihood solution gives the 
following LF parameters:
 
\begin{equation}
\phi(L) \propto \left\{\!
\begin{array}{ll}
L^{-1.52}; & L<10^{52.5}{\rm\, erg\,s^{-1}}; \\
L^{-2.00}; & L>10^{52.5}{\rm\, erg\,s^{-1}}.
\end{array}\right.
\end{equation}

\section{DISCUSSION AND IMPLICATIONS}

\label{dis.sec}

Figure~\ref{zdist.fig} shows the cumulative redshift distribution of the
largest homogeneous and unbiased sample of GRBs to date. It contains redshift 
information on all the 69 TOUGH bursts, including limits. The conservative 
systematic error region (hatched) of the TOUGH sample redshift distribution is 
significantly smaller than for previous samples 
\citep[e.g. Figure~2 in][]{palliAIP}. This allows the rejection of various model 
predictions (e.g.\ compare the dashed and dotted curves in 
Fig.~\ref{zdist.fig}). Thus, we have been able to confirm previous findings 
\citep[e.g.][]{kistler09,virgili11,robellis} that the GRB rate at high redshift 
($z \gtrsim 3$) appears to be in excess of predictions based on the 
assumption that it should follow conventional determinations of the 
star-formation history of the universe, combined with an estimate of its 
likely metallicity dependence \citep{langer}.  
\par
It is possible that star formation at high redshifts has been significantly 
underestimated. Even at $z \sim 2$ it appears that the galaxy LF has a 
substantially steeper faint-end slope than locally \citep[e.g.][]{reddy09}, 
while recent LF studies in the Hubble Ultra-Deep Field have concluded that 
at $z \gtrsim 7$ so-far undetected galaxies are likely to completely dominate 
the total star formation activity \citep[e.g.][]{bouwens11,nial12}. This 
picture is supported by previous observations of damped Ly$\alpha$ absorbers 
\citep[e.g.][]{johan99,haehnelt,joop} and Ly$\alpha$ emitting galaxies 
\citep[e.g.][]{johan03,palli05}, as well as by recent simulation studies 
\citep{choi}. Alternatively, it could be that GRB production is substantially 
enhanced in the conditions of early star formation, beyond the 
metallicity-dependent rate correction already applied. In the long run, 
large complete samples of GRB redshifts should shed light on whether the GRB 
rate is proportional to SFR or whether other effects play an important role. 
\par
The sampling error and the conservative systematic error region are shown
separately to clearly illustrate that incompleteness dominates the sample,
and more is gained by reducing the systematics rather than increasing the 
sample size. Using both error regions we can set a conservative limit on the 
maximum number of \emph{Swift} bursts at $z > 6$ ($z > 7$): 14\% (5\%). This 
is fully consistent with the models which predict between 2 (\emph{SFR1}) 
and 13 (\emph{SFR2}) bursts per year (all sky) at $6 < z < 9$ to 
\emph{Swift}/BAT limits.
\par
\begin{figure}
\epsscale{1.15}
\plotone{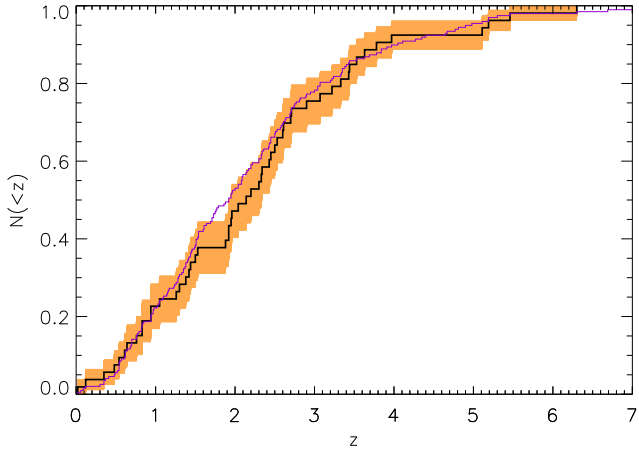}
\caption{Thick solid curve and shaded region are defined as in  
Fig.~\ref{zdist.fig} caption. The thin solid curve is the cumulative fraction 
of GRBs as a function of redshift for all long \emph{Swift} bursts to date 
(2012 April 1: 198 redshifts) with $\langle z \rangle = 2.16$.}
\label{zall.fig}
\end{figure}
The average (median) redshift of the 53 TOUGH bursts is $\langle z \rangle 
= 2.23$ ($\tilde{z}  = 2.14$), significantly lower than the early 
\emph{Swift} results indicated \citep[e.g.][]{palli28}. This difference may 
simply reflect the comparatively small samples analyzed in that paper, but 
could also be due to an increased success in measuring redshifts $z < 2$ using 
weaker absorption lines in afterglow spectra, and via host galaxies. The mean 
redshift could be as low as $\langle z \rangle \sim 1.7$ (upper boundary of 
the hatched region) although it is unlikely that the majority of bursts with 
unknown redshifts would be located at very small distances. In fact, it is 
more probable that $\langle z \rangle \gtrsim 2.20$ since we have only targeted
the brightest galaxies in the sample ($R \lesssim 25$\,mag) for spectroscopic 
follow-up. In Fig.~\ref{zall.fig}, we compare the TOUGH redshift distribution 
to the one obtained from all long-duration \emph{Swift} bursts. The agreement 
is good in general but the TOUGH sample includes relatively more bursts in the 
range $1.8 \lesssim z \lesssim 2.5$. This can be explained by the fact that 
Ly$\alpha$ is located in the blue end of the optical at this redshift range
which makes it easy to miss a damped Ly$\alpha$ absorber. On the other hand,
emission lines are readily detectable with X-shooter at these redshifts.
\par
As discussed in \citet{jensLP}, one of the TOUGH selection criteria (requiring
the XRT error circle radius to be less than 2\arcsec) might slightly bias
our sample against faint, hence potentially high-redshift, bursts. In total, 
only two events were rejected due to a large error radius (GRBs\,050412 and 
061102). In both cases, the X-ray afterglows were unusually faint, so this is 
an estimate of the fraction\footnote{The total number of bursts here being the 
TOUGH 69 along with those two with a large error radius.} (2/71) of 
\emph{Swift} GRBs that would fail the TOUGH positional accuracy selection 
criterion which might well be at high redshift. Further to this, a comparable 
bias is introduced by the exclusion of bursts without a detected X-ray 
afterglow at all (5/180).\footnote{\emph{Swift} detected 180 long bursts 
during the TOUGH time period in which the XRT was quickly ($<1$\,hr) 
repointed.} All five GRBs are low-significance detections and are thus likely
to be spurious. In total, a maximum of 5.5\% of \emph{Swift} long bursts could 
be at high redshift and not enter the TOUGH sample.
\par
We have now reached a point in GRB research where a single burst rarely 
elucidates and illuminates our general understanding of the field. It is 
important to focus on well-defined samples and population studies, where 
systematics and biases can be minimized. \emph{Swift} has made it possible to 
build such a sample and thanks to new available instrumentation, such as the 
VLT/X-shooter \citep{vernet}, we can continue to follow this track into the 
future.


\acknowledgments

We thank the referee for providing excellent comments in a timely manner and 
Andrew J. Levan for providing us with the NIR afterglow image for 
GRB\,060814. We also thank Thomas Kr\"uhler, Dan Perley, and Ruben Salvaterra 
for very informative discussions. Based on observations collected at the 
European Southern Observatory, Chile, as part of the large program 177.A-0591. 
P.J. acknowledges support by a Project Grant from the Icelandic Research Fund. 
D.M. acknowledges financial support from the Instrument Center for Danish 
Astrophysics. The Dark Cosmology Centre is funded by the Danish National 
Research Foundation. This work has made use of the University of Hertfordshire 
Science and Technology Research Institute high-performance computing facility.

\clearpage


\begin{thebibliography}{}


\bibitem[Appenzeller et al.(1998)]{appen} 
Appenzeller, I., et al. 1998, The Messenger 94, 1

\bibitem[Bloom et al.(2002)]{josh} 
Bloom, J. S., Kulkarni, S. R., \& Djorgovski, S. G.
2002, AJ, 123, 1111

\bibitem[Bouwens et al.(2008)]{bouwens} 
Bouwens, R. J., Illingworth, G. D., Franx, M., \& Ford, H.
2008, ApJ, 686, 230

\bibitem[Bouwens et al.(2011)]{bouwens11} 
Bouwens, R. J., et al. 2011, ApJ, in press (arXiv:1105.2038)

\bibitem[Butler et al.(2010)]{butler} 
Butler, N. R., Bloom, J. S., \& Poznanski, D. 2010, ApJ, 711, 495

\bibitem[Cao et al.(2011)]{cao} 
Cao, X.-F., Yu, Y.-W., Cheng, K.~S., \& Zheng, X.-P. 
2011, MNRAS, in press (arXiv:1101.0866)

\bibitem[Campana et al.(2010)]{campana} 
Campana, S., Th\"one, C. C., de Ugarte Postigo, A., Tagliaferri, G., 
Moretti, A., \& Covino, S. 2010, MNRAS, 402, 2429

\bibitem[Campisi et al.(2010)]{maria} 
Campisi, M. A., Li, L.-X., \& Jakobsson, P.
2010, MNRAS, 407, 1972

\bibitem[Chapman et al.(2007)]{bob} 
Chapman, R., Tanvir, N. R., Priddey, R. S., \& Levan, A. J. 
2007, MNRAS, 382, L21

\bibitem[Choi \& Nagamine(2011)]{choi} 
Choi, J.-H., \& Nagamine, K. 2012, MNRAS, 419, 1280

\bibitem[Christensen et al.(2004)]{lise} 
Christensen, L., Hjorth, J., \& Gorosabel, J.
2004, A\&A, 425, 913

\bibitem[Cobb et al.(2006)]{cobb} 
Cobb, B. E., Bailyn, C. D., van Dokkum, P. G., \& Natarajan, P.
2006, ApJ, 645, L113

\bibitem[Courbin et al.(2000)]{courbin}
Courbin, F., Magain, P., Kirkove, M., \& Sohy, S. 2000, ApJ, 529, 1136

\bibitem[Coward(2005)]{coward05}
Coward, D. M. 2005, MNRAS, 360, L77

\bibitem[Coward et al.(2008)]{coward08}
Coward, D. M., Guetta, D., Burman, R. R., \& Imerito, A. 
2008, MNRAS, 386, 111

\bibitem[Daigne et al.(2006)]{daigne} 
Daigne, F., Rossi, E. M., \& Mochkovitch, R.
2006, MNRAS, 372, 1034

\bibitem[Dall'Aglio et al.(2008)]{dall} 
Dall'Aglio, A., Wisotzki, L., \& Worseck, G.
2008, A\&A, 491, 465

\bibitem[Dong \& Lu(2009)]{dong} 
Dong, Y.-M., \& Lu, T. 2009, RAA, 9, 95


\bibitem[Foley et al.(2008)]{foley}
Foley, S., McGlynn, S., Hanlon, L., McBreen, S., \& McBreen, B.
2008, A\&A, 484, 143

\bibitem[Fruchter et al.(2006)]{andy}
Fruchter, A. S., et al. 2006, Nature, 441, 463

\bibitem[Fynbo et al.(1999)]{johan99}
Fynbo, J. P. U., M\o ller, P., \& Warren, S. J.
1999, MNRAS, 305, 849

\bibitem[Fynbo et al.(2002)]{johan02}
Fynbo, J. P. U., et al. 2002, A\&A, 388, 425

\bibitem[Fynbo et al.(2003)]{johan03}
Fynbo, J. P. U., Ledoux, C., M\o ller, P., Thomsen, B., \& Burud, I.
2003, A\&A, 407, 147

\bibitem[Fynbo et al.(2008)]{johan}
Fynbo, J. P. U., Prochaska, J. X., Sommer-Larsen, J.,
Dessauges-Zavadsky, M., \& M\o ller, P. 2008, ApJ, 683, 321

\bibitem[Fynbo et al.(2009)]{johan_sample}
Fynbo, J. P. U., et al. 2009, ApJS, 185, 526 

\bibitem[Gehrels et al.(2004)]{gehrels}
Gehrels, N., et al. 2004, ApJ, 611, 1005

\bibitem[Gorosabel et al.(2004)]{javier}
Gorosabel, J., Lund, N., Brandt, S., Westergaard, N. J., \& 
Castro Cer\'on, J. M. 2004, A\&A, 427, 87

\bibitem[Grupe et al.(2007)]{grupe}
Grupe, D., et al. 2007, AJ, 133, 2216

\bibitem[Guetta et al.(2005)]{guetta05}
Guetta, D., Piran, T., \& Waxman, E. 2005, ApJ, 619, 412


\bibitem[Guetta \& Piran(2007)]{guetta07}
Guetta, D., \& Piran, T. 2007, JCAP, 7, 3

\bibitem[Haehnelt et al.(2000)]{haehnelt}
Haehnelt, M. G., Steinmetz, M., \& Rauch, M.
2000, ApJ, 534, 594

\bibitem[Hjorth et al.(2003)]{jens} 
Hjorth, J., et al. 2003, Nature, 423, 847

\bibitem[Hjorth et al.(2012)]{jensLP} 
Hjorth, J., et al. 2012, ApJ, submitted

\bibitem[Hopkins \& Beacom(2006)]{hop} 
Hopkins, A. M., \& Beacom, J. F. 2006, ApJ, 651, 142

\bibitem[Jakobsson et al.(2004)]{pallidark}
Jakobsson, P., Hjorth, J., Fynbo, J. P. U., Watson, D., Pedersen, K., 
Bj\"ornsson, G., \& Gorosabel, J. 2004, ApJ, 617, L21

\bibitem[Jakobsson et al.(2005)]{palli05} 
Jakobsson, P., et al. 2005, MNRAS, 362, 245

\bibitem[Jakobsson et al.(2006)]{palli28} 
Jakobsson, P., et al. 2006, A\&A, 447, 897

\bibitem[Jakobs\-son et al.(2009)]{palliAIP}
Jakobsson, P., Malesani, D., Fynbo, J. P. U., Hjorth, J., \& 
Milvang-Jensen, B. 2009, in AIP Conference Proceedings 1133,
Gamma-Ray Bursts: Sixth Huntsville Symposium, ed. C. Meegan, 
N. Gehrels, \& C. Kouveliotou (New York: AIP), 455

\bibitem[Jakobs\-son et al.(2011a)]{palliAdSpR}
Jakobsson, P., Malesani, D., Hjorth, J., Fynbo, J. P. U., \&
Milvang-Jensen, B. 2011a, AdSpR, 47, 1416

\bibitem[Jakobs\-son et al.(2011b)]{palliAN}
Jakobsson, P., Malesani, D., Hjorth, J., Fynbo, J. P. U., \&
Milvang-Jensen, B. 2011b, AN, 332, 276

\bibitem[Jakobs\-son et al.(2011c)]{palliUSA}
Jakobsson, P., Malesani, D., Hjorth, J., Fynbo, J. P. U., \&
Milvang-Jensen, B. 2011c, in AIP Conference Proceedings 1358,
Gamma Ray Bursts 2010, ed. J. E. McEnery, J. L. Racusin \& 
N. Gehrels (New York: AIP), 265

\bibitem[Jaunsen et al.(2008)]{andreas} 
Jaunsen, A. O., et al. 2008, ApJ, 681, 453

\bibitem[Jarosik et al.(2011)]{jarosik} 
Jarosik, N., et al. 2011, ApJS, 192, 14

\bibitem[Kistler et al.(2008)]{kistler} 
Kistler, M. D., Y\"uksel, H., Beacom, J. F., \& Stanek, K. Z.
2008, ApJ, 673, L119

\bibitem[Kistler et al.(2009)]{kistler09} 
Kistler, M. D., Y\"uksel, H., Beacom, J. F., Hopkins, A. M., 
\& Wyithe, J. S. B. 2009, ApJ, 705, L104

\bibitem[Kr\"uhler et al.(2012)]{thomas} 
Kr\"uhler, T., et al. 2012, ApJ, submitted

\bibitem[Langer \& Norman(2006)]{langer}
Langer, N., \& Norman, C. A. 2006, ApJ, 638, L63

\bibitem[Le \& Dermer(2007)]{le} 
Le, T., \& Dermer, C. D. 2007, ApJ, 661, 394

\bibitem[Le Floc'h et al.(2003)]{lefl} 
Le Floc'h, E., et al. 2003, A\&A, 400, 499

\bibitem[Levan et al.(2006)]{andrew} 
Levan, A. J., Tanvir, N. R., Rol, E., Fruchter, A., \& Adamson, A.
2006, GCN Circ. 5455

\bibitem[Letawe et al.(2008)]{letawe08} 
Letawe, Y., Magain, P., Letawe, G., Courbin, F., \& Hutsem\'ekers, D.
2008, ApJ, 679, 967

\bibitem[Levesque et al.(2010)]{levesque}
Levesque, E. M., Kewley, L. J., Berger, E., \& Jabran Zahid, H.
2010, AJ, 140, 1557

\bibitem[Li(2008)]{li} 
Li, L.-X. 2008, MNRAS, 388, 1487

\bibitem[Liang et al.(2007)]{liang}
Liang, E., Zhang, B., Virgili, F., \& Dai, Z. G. 2007, ApJ, 662, 1111

\bibitem[Malesani(2006)]{dm814} 
Malesani, D. 2006, GCN Circ. 5456

\bibitem[Malesani et al.(2012)]{danieleLP} 
Malesani, D., et al. 2012, ApJ, in preparation


\bibitem[Milvang-Jensen et al.(2012)]{boLP} 
Milvang-Jensen, B., et al. 2012, ApJ, submitted

\bibitem[Natarajan et al.(2005)]{priya} 
Natarajan, P., Albanna, B., Hjorth, J., Ramirez-Ruiz, E., Tanvir, N.,
\& Wijers, R. 2005, MNRAS, 364, L8

\bibitem[Ovaldsen et al.(2007)]{ovald} 
Ovaldsen, J.-E., et al. 2007, ApJ, 662, 294

\bibitem[Paczy\'nski(1998)]{pacz} 
Paczy\'nski, B. 1998, ApJ, 494, L45

\bibitem[Perley et al.(2009)]{perley} 
Perley, D. A., et al. 2009, AJ, 138, 1690


\bibitem[Porciani \& Madau(2001)]{porc} 
Porciani, C., \& Madau, P. 2001, ApJ, 548, 522

\bibitem[Predehl \& Schmitt(1995)]{pre}
Predehl, P., \& Schmitt, J. H. M. M. 1995, A\&A, 293, 889

\bibitem[Reddy \& Steidel(2009)]{reddy09} 
Reddy, N. A., \& Steidel, C. C. 2009, ApJ, 692, 778

\bibitem[Reddy et al.(2008)]{reddy} 
Reddy, N. A., Steidel, C. C., Pettini, M., Adelberger, K. L., Shapley, A. E., 
Erb, D. K., \& Dickinson, M. 2008, ApJS, 175, 48

\bibitem[Robertson \& Ellis(2012)]{robellis}
Robertson, B. E., \& Ellis, R. S. 2012, ApJ, 744, 95

\bibitem[Rol et al.(2006)]{rol}
Rol, E., Jakobsson, P., Tanvir, N., \& Levan, A. 2006, GCN Circ. 5555


\bibitem[Salvaterra \& Chincarini(2007)]{salvaKING} 
Salvaterra, R., \& Chincarini, G. 2007, ApJ, 656, L49

\bibitem[Salvaterra et al.(2007)]{salvaZ} 
Salvaterra, R., Campana, S., Chincarini, G., Tagliaferri, G., \& Covino, S.
2007, MNRAS, 380, L45

\bibitem[Salvaterra et al.(2009)]{salva} 
Salvaterra, R., Guidorzi, C., Campana, S., Chincarini, G., \& Tagliaferri, G.
2009, MNRAS, 396, 299

\bibitem[Salvaterra et al.(2012)]{ruben2012}
Salvaterra, R., et al. 2012, ApJ, 749, 68

\bibitem[Savaglio et al.(2009)]{savaglio} 
Savaglio, S., Glazebrook, K., \& Le Borgne, D. 2009, ApJ, 691, 182

\bibitem[Schaye(2001)]{joop} 
Schaye, J. 2001, ApJ, 559, L1


\bibitem[Schmidt(2009)]{schmidt09} 
Schmidt, M. 2009, ApJ, 700, 641

\bibitem[Soderberg et al.(2006)]{sod} 
Soderberg, A. M., et al. 2006, Nature, 442, 1014

\bibitem[Songaila(2004)]{son}
Songaila, A. 2004, AJ, 127, 2598


\bibitem[Tanvir \& Jakobsson(2007)]{np} 
Tanvir, N. R., \& Jakobsson, P. 2007, RSPTA, 365, 1377

\bibitem[Tanvir et al.(2008)]{nial} 
Tanvir, N. R., et al. 2008, MNRAS, 388, 1743

\bibitem[Tanvir et al.(2012)]{nial12} 
Tanvir, N. R., et al. 2012, ApJ, in press (arXiv:1201.6074)

\bibitem[Th\"one et al.(2007)]{ct} 
Th\"one, C. C., Perley, D. A., \& Bloom, J. S. 2007, GCN Circ. 6663

\bibitem[Vernet et al.(2011)]{vernet}
Vernet, J., et al. 2011, A\&A, 536, 105

\bibitem[Virgili et al.(2009)]{virgili}
Virgili, F. J., Liang, E.-W., \& Zhang, B. 2009, MNRAS, 392, 91

\bibitem[Virgili et al.(2011)]{virgili11}
Virgili, F. J., Zhang, B., Nagamine, K., \& Choi, J.-H.
2011, MNRAS, 417, 3025

\bibitem[van Dokkum(2001)]{vanD}
van Dokkum, P. G. 2001, PASP, 113, 1420

\bibitem[Wanderman \& Piran(2010)]{wand} 
Wanderman, D. \& Piran, T. 2010, MNRAS, 406, 1944

\bibitem[Wijers et al.(1998)]{ralph} 
Wijers, R. A. M. J., Bloom, J. S., Bagla, J. S., \& Natarajan, P.
1998, MNRAS, 294, L13

\bibitem[Woosley \& Bloom(2006)]{WB} 
Woosley, S. E., \& Bloom, J. S. 2006, ARA\&A, 44, 507

\bibitem[Zitouni et al.(2008)]{zitouni} 
Zitouni, H., Daigne, F., Mochkovich, R., \& Zerguini, T. H.
2008, MNRAS, 386, 1597

\end{thebibliography}
\end{document}